\documentclass[%
 reprint,
superscriptaddress,
 amsmath,amssymb,
 aps,
 pra,
]{revtex4-1}

\usepackage{graphicx}
\usepackage{dcolumn}
\usepackage{bm}
\usepackage{upgreek} 
\usepackage{color} 
\usepackage{hyperref}

\begin{document}

\preprint{APS/123-QED}

\title{Loading and spatially-resolved characterization of a cold atomic ensemble inside a hollow-core fiber}

\author{Thorsten Peters}
\affiliation{Institute of Applied Physics, Technical University of Darmstadt, Hochschulstrasse 6, 64289 Darmstadt, Germany}
\author{Leonid P. Yatsenko}
\affiliation{Institute of Physics, National Academy of Science of Ukraine, Nauky Ave. 46, Kyiv 03028, Ukraine}
\author{Thomas Halfmann}
\affiliation{Institute of Applied Physics, Technical University of Darmstadt, Hochschulstrasse 6, 64289 Darmstadt, Germany}

\date{\today}

\begin{abstract}
We present a thorough experimental investigation of the loading process of laser-cooled atoms from a magneto-optical trap into an optical dipole trap located inside a hollow-core photonic bandgap fiber, followed by propagation of the atoms therein. This, e.g., serves to identify limits to the loading efficiency and thus optical depth which is a key parameter for applications in quantum information technology.
Although only limited access in 1D is available to probe atoms inside such a fiber, 
we demonstrate that a detailed spatially-resolved characterization of the loading and trapping process 
along the fiber axis is possible by appropriate modification of probing techniques combined with theoretical analysis. 
Specifically, we demonstrate the loading of up to $2.1 \times 10^5$ atoms with a transfer efficiency of $2.1~\%$ during the course of $50$~ms and a peak loading rate of $4.7 \times 10^3$ atoms ms$^{-1}$ resulting in a peak atomic number density on the order of $10^{12}~$cm$^{-3}$. Furthermore, we determine the evolution of the spatial density (profile) and ensemble temperature as it approaches its steady-state value of $T=1400~\upmu$K, as well as loss rates, axial velocity and acceleration. 
The spatial resolution along the fiber axis reaches a few millimeters, which is much smaller than the typical fiber length in experiments. 
We compare our results to other fiber-based as well as free-space optical dipole traps and discuss the potential for further improvements.

\end{abstract}

\maketitle


\section{\label{sec:introduction}Introduction}

Since its proposal by Ol'shanii \textit{et al.} \cite{OOL93} laser guiding of atoms into or through hollow-core optical fibers \cite{DAV19} has received increasing experimental interest over the past years \cite{RMV95,RDC96,INS96,RZD97,DHB99,MCA00,CWS08,BHB09,BHP11,TK07,VMW10,PBH12,BHP14,LNK17,OTB14,LWK18,HPL19,HPB18,XLC18,NLW18,XLC19,YB19,YDF19,OYV19,HLL20,WCX20,LXH20,LIW20,LXC20}
This is due to their potential for controlled guiding of atoms over macroscopic distances for atom interferometry \cite{VMW10,XLC18} and quantum optics \cite{BHB09,NA17}. The latter application requires long coherence time \cite{XLC19} and strong light-matter coupling, i.e., transverse tightly confined atomic ensembles of large optical depth $d_{opt}=\sigma_0 n L$ \cite{BHP14}, where $\sigma_0$ is the atomic absorption cross-section, $n$ is the atomic number density, and $L$ is the length of the ensemble. Therefore, efforts have been devoted in recent years to load laser-cooled atoms from a magneto-optical trap (MOT) located at some distance from the fiber into their hollow core \cite{CWS08,BHB09,BHP11,TK07,VMW10,PBH12,BHP14,OTB14,LNK17,LWK18,HPB18,HPL19,NLW18,XLC18,XLC19,YB19,YDF19,OYV19,HLL20,WCX20,LXH20,LIW20,LXC20} by use of a standing- or traveling-wave optical dipole trap \cite{GWO00}.

After pioneering work on laser-guiding of atoms through hollow fibers which had to rely on lossy glass capillaries of large core diameters \cite{RMV95,RDC96,INS96,RZD97,DHB99,MCA00}, later experiments employed single-mode, low-loss photonic cyrstal fibers \cite{R03} due to their more advantageous properties. Nowadays, basically two types of fibers are applied. Kagome-structured fibers \cite{CBL06} and hollow-core photonic bandgap fibers (HCPBGFs) \cite{CMK99}. The first type of fiber provides single-mode guidance of light over a large wavelength range, low birefringence, and low numerical aperture for core diameters of several ten micrometers. The large bandwidth and low birefringence make Kagome fibers ideally suited for far-off-resonance guiding with on-resonance probing and controlled manipulation of different atomic species. Their low numerical aperture, and thus slowly diverging laser beams, allow for direct transport from the distant MOT into the fiber core \cite{OTB14,HPB18,LWK18,XLC18,HPL19,HLL20,WCX20,LXH20,LIW20}. Experimental and theoretical loading studies using Kagome fibers were reported recently \cite{LWK18,HPB18,WCX20}. A drawback of these type of fibers is their relatively large core diameter, resulting in weaker light-matter coupling strengths, compared to HCPBGFs.
HCPBGFs, on the other hand, are available with core diameters below $10~\upmu$m to enable stronger light-matter coupling and suffer less from micro-lensing \cite{NLW18,SPW19}, but provide smaller guiding bandgaps and larger, uncontrolled birefringence \cite{WLG05}. Nonetheless, efficient loading of atoms into such fibers \cite{CWS08,BHB09,BHP14} and quantum optics experiments \cite{BHB09,BSH16}, even down to the quantum level \cite{PWN20}, were demonstrated recently. The larger numerical aperture of HCPBGFs usually prohibits a direct transfer of atoms from the MOT into the fiber, as the trapping potential (proportional to the laser intensity) quickly decreases outside of the fiber. Thus, a different approach to loading has to be chosen. 

So far, only loading via free-fall, magnetic guiding, and hollow-beam guiding was investigated experimentally by Bajcsy \textit{et al.} \cite{BHP11} for a HCPBGF. The more efficient loading technique, using a dark funnel to guide the atoms into the fiber \cite{BHP14,BSH16,PWN20}, has however not yet been investigated thoroughly. Furthermore, spatially-resolved detection within such fiber has not been demonstrated so far. Such information is however important for optimizing the loading efficiency and minimizing losses within the fiber.
We report in the following on a detailed experimental study of loading atoms from a MOT into a far-off-resonance trap (FORT) \cite{CBA86,MCH93a} located inside a HCPBGF by using a dark funnel guide. Although FORTs in free space have been studied for years (see, e.g., \cite{MCH93a,KCM00,GWO00}) and a variety of probing techniques are available, there are some important differences compared to a FORT located inside a hollow-core fiber. Loading and probing is only possible along the direction of the fiber axis and the axial extension of the FORT can be four orders of magnitude larger than the radial extension. The atoms are therefore basically untrapped along the fiber axis and propagation of the ensemble becomes relevant. We will present experimental data along with comparison to theoretical models to characterize the system--even spatially-resolved along the fiber axis. Specifically, we determine the loading efficiency into the fiber, and loss rates, temperature, velocity as well as the spatial density distribution inside the fiber. This allows us to identify limits to the loading efficiency and possible improvements as well as to compare our loading scheme to others.

This paper is organized as follows: In Sec.~\ref{sec:experimental_details} we give an overview of the experimental system and describe the probing techniques applied in our work. 
In Sec.~\ref{sec:spatial} we study the atomic ensemble inside the fiber with a focus on spatially-resolved measurements in both radial and axial dimensions. Next, we study the number and propagation of atoms loaded from the MOT into the FORT for various parameters with a focus on efficient loading in Sec.~\ref{sec:atom_number}. Finally, we present a summary and outlook in Sec.~\ref{sec:summary}.

\section{\label{sec:experimental_details}Experimental details}
Details on the experimental setup and sequence can be found in our previous work using a HCPBGF \cite{BHP14,BSH16,PWN20}. 
Thus, we here only briefly summarize the setup.

\subsection{\label{sec:experimental_setup}Experimental setup and sequence}
\begin{figure}[b]
\includegraphics[width=8.5cm]{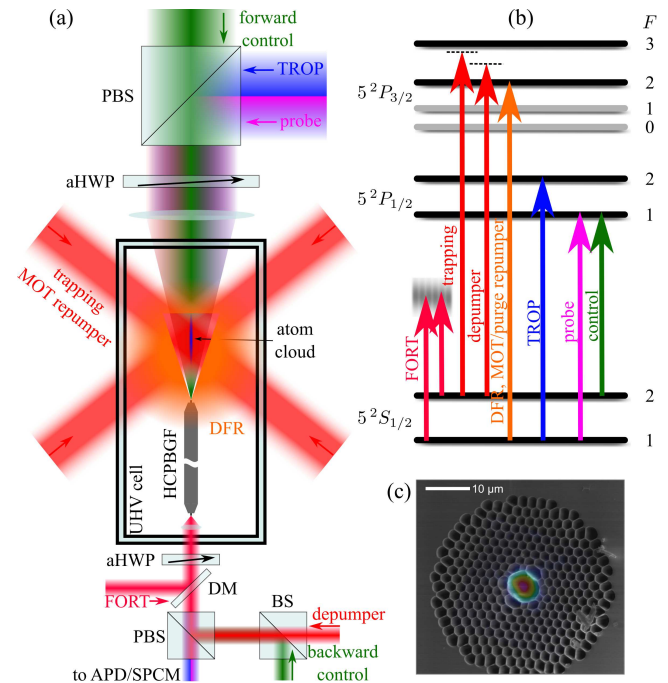}
\caption{\label{fig:setup} (a) Schematic experimental setup. DM: dichroic mirror; aHWP: achromatic half wave plate; PBS: polarizing beam splitter: BS: beam splitter; APD: avalanche photodiode; SPCM: single-photon counting module. (b) Coupling scheme. (c) Scanning electron micrograph of the HCPBGF and false color intensity profile of the FORT.}
\end{figure} 
We first load laser-cooled $^{87}$Rb atoms into a cigar-shaped vapor cell MOT located about $5.5$~mm above the tip of the vertically mounted HCPBGF (\textit{NKT Photonics} HC800-02) (see Fig.~\ref{fig:setup}). The HCPBGF has a length of $14$~cm and a core diameter of $d_c \sim 7~\upmu$m. The $1/e$ intensity mode field diameter inside the fiber is around $5.5~\upmu$m. After the MOT is loaded for about 1~s yielding typically around $10^7$ atoms, we shift the cloud down towards the HCPBGF tip during the next $40$~ms by applying a magnetic offset field in the vertical direction. Simultaneously, we transfer the atoms from the bright MOT into a \textit{dark funnel} guide \cite{BHP14} by switching off the MOT repumper and ramping up the quadrupole gradient of the MOT coils which compresses the atom cloud towards the HCPBGF. Also, the trapping detuning is increased to $-30~$MHz for optimized cooling during the transfer. The dark funnel guide consists of two nearly orthogonally aligned repumper beams with a funnel-shaped shadow in their center, termed dark funnel repumper (DFR) in the following. This dark funnel is aligned from the side onto the HCPBGF. In order to improve the density above the HCPBGF and confine the atoms in $F=1$ we use a depumper beam traveling upwards through the HCPBGF \cite{APE94}. The depumper is locked to the cross-over transition  $5 ^2S_{1/2}\,F=2 \rightarrow 5 ^2P_{3/2}\, F'=2,3$. As the effective transition light-shift by the FORT is estimated to be near $+130~$MHz for our typical FORT depth, the depumper is almost resonant within and slightly above the HCPBGF, where loading and guiding is relevant. All trapping, MOT repumper, and DFR beams have diameters of nearly 20~mm thus covering the whole region of MOT and HCPBGF tip where trapping and guiding is relevant.
After the cloud has reached the fiber tip we hold it for $10~$ms in place. Here, the atoms are transferred into a FORT of depth $T_F \lesssim 4.3~$mK. The FORT system consists of two independent unstabilized laser diodes operating near 820~nm and of orthogonal linear polarizations, combined at a polarizing beam splitter. Using a FORT system operating at multiple frequencies was shown to improve the loading efficiency into the HCPBGF as compared to single-mode operation \cite{BHP14}. As the FORT beam quickly diverges above the HCPBGF, its trapping potential decreases quickly with the square of the distance from the fiber tip. A simulation of the spatial trapping potential shows that the atoms have to be brought within a distance of around $200~\upmu$m of the fiber tip for the FORT being strong enough to guide the atoms into the core. Thus, cooling and compression of the atom cloud above the HCPBGF is crucial for obtaining a good loading efficiency. Once the atoms are inside the fiber they can be probed as described in the next section. In order to prevent detection of atoms that might still be left above the HCPBGF during probing, we use a strong repumper beam above the fiber to purge the volume through which the probe beam propagates. 
This confines the population in $F=2$ above the HCPBGF, whereas we always probe the atoms in $F=1$ inside the fiber. Except for the repumper, all laser beams are temporally modulated by acousto-optic modulators (AOMs).

\subsection{\label{sec:techniques}Probing techniques}

\subsubsection{\label{sec:techniques_OP}Time-resolved optical pumping}
The number of atoms inside the HCPBGF, $N_{a}$, can be inferred, e.g., by nonlinear saturation measurements \cite{BHP11} or spectroscopic studies followed by comparison to a simulation \cite{HPB18}. A simpler method infers $N_{a}$ from time-resolved optical pumping (TROP) \cite{BHP14}.

\begin{figure}[tb]
\includegraphics[width=8.5cm]{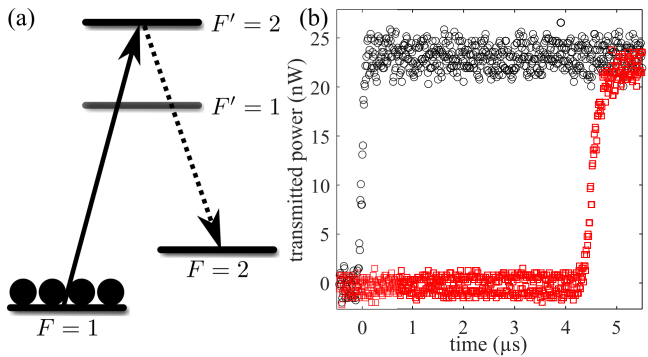}
\caption{\label{fig:TROP} (a) Level scheme for TROP on the D$_1$ line. (b) Transmission for TROP  with $N_{a}=2.1(2)\times 10^5$ atoms (\textcolor{red}{{\tiny $\square$}}) and without atoms ($\circ$) inside the HCPBGF. The vertical axis is calibrated to show the actual power inside the HCPBGF.}
\end{figure}

Here, population is first prepared in the lower ground state $F=1$. Then, the optical trap is quickly switched off and a strong pump laser, tuned, e.g., to the transition $F=1 \rightarrow F'=2$ [see Fig.~\ref{fig:TROP}(a)], 
is sent through the fiber and detected by an avalanche photodiode (APD). This beam optically pumps the population into $F=2$ where it does not interact with the pump anymore. The medium thus becomes transparent. If the APD is calibrated and the transmission from the fiber to the APD is known, the difference of the temporally-resolved transmissions without and with atoms is thus a measure for the energy absorbed by the atoms until they are in the uncoupled level $F=2$. With the branching ratio from  $F'=2$ into $F=1,2$, here 0.5, the number of photons an atom absorbs on the average before it ends up in $F=2$ is 2. From the known energy of a photon, $hc/\lambda$, we therefore get $N_a$.
The result of such measurement is shown in Fig.~\ref{fig:TROP}(b). We use this technique to study the loading efficiency for various parameters as we will discuss in the following sections.

\subsubsection{\label{sec:techniques_EIT}Electromagnetically induced transparency}
The other technique suitable to extract information from within the HCPBGF is electromagnetically induced transparency (EIT) \cite{H97,FIM05}. Here, a weak probe and a strong control field couple two long-lived states $|1,2\rangle$ via a decaying intermediate state $|3\rangle$, e.g., in a $\Lambda-$scheme. The strong control field renders the previously opaque medium transparent for the weak probe within a frequency window around two-photon resonance between the two long-lived states. The width of the EIT window $\Delta\omega_{EIT}=\Omega_c^2/(\Gamma\sqrt{d_{opt}})$ depends on the control Rabi frequency $\Omega_c$, the intermediate state decay rate $\Gamma$ and the optical depth $d_{opt}$ of the atomic ensemble. The depth of the EIT window depends on the ground state dephasing rate $\gamma_{21}$ between states $|1,2\rangle$ relative to $\Delta\omega_{EIT}$. This enables determination of $\gamma_{21}$ from EIT spectra \cite{PLS12}. For $\Lambda-$type coupling of nearly energetically degenerate ground states and collinear probe and control beams with wavevectors $\bm{k}_{p,c}$ the two-photon resonance is not very susceptible to atomic motion, as the corresponding Doppler shift $\left(\bm{k}_p - \bm{k}_c\right)\cdot \bm{v}_{a}$ is very small, where $\bm{v}_{a}$ is the atomic velocity along the laser beam direction. This allows for determining the effective dephasing rate $\gamma_{21}$ comprised of transit-time broadening, two-photon linewidth, and inhomogeneous transition shifts due to the radially varying control field intensity \cite{BSH16}. For simplicity we do not discriminate here between reversible dephasing processes, e.g., due to a spatial magnetic field gradient and irreversible decoherence, e.g., due to transit-time broadening.  For anti-collinear probe and control beams, however, the two-photon resonance is very susceptible to atomic motion as the two-photon Doppler shifts is now  $\left(\bm{k}_p - \bm{k}_c\right)\cdot \bm{v}_{a}\approx 2\cdot \bm{k}_{p,c}\cdot \bm{v}_{a}$. This can be used in ultracold media to obtain information on the atomic velocity (distribution) along the laser beam direction \cite{PWB12}.
Thus, EIT can be used to obtain information on the magnetic field (gradient) and atomic velocities inside the HCPBGF, depending on the configuration used.

\section{\label{sec:experimental}Experimental studies}

In the following experimental studies we demonstrate how to gain information on the atoms as they propagate through the HCPBGF. This allows us to identify limits to the loading efficiency, loss rates, and coherence time, which are important parameters for applications. Unlike in free-space setups, optical access from the side is not possible for a hollow-core fiber. Therefore, information on the number, position and velocity of the atoms, temperature, radial distribution, and magnetic field (gradient) have to be obtained from measurements performed along the fiber axis only. A detailed characterization of the FORT potential can be found in App.~\ref{sec:appendix_FORT}, as it is required for a quantitative analysis of the following results.

\subsection{\label{sec:spatial} Spatially-resolved characterization of the atomic ensemble}
Typically, HCPBGFs are loaded with the aim to maximize the number of atoms inside the fiber as to create the highest possible $d_{opt}$ or to guide atoms with losses from the FORT as small as possible. This is achieved when the fiber can be filled over a length as large as possible. However, experimental studies of the confined atomic ensemble thereafter only yield an average over the ensemble length of several centimeters (up to the length of the HCPBGF; 14~cm in our case). This impedes gaining information on, e.g., possible magnetic field gradients along the fiber, the axial atomic density profile etc. Also the radial distribution of atoms is of interest. It relates to the temperature of the atoms inside the FORT and affects inhomogeneous broadening as the atomic density profile width is typically comparable to the mode field diameter of the laser beams.
We thus describe in the following methods to extract such information. To this end, we load the HCPBGF only with a short pulse of atoms and interrogate them as a function of time $t_{if}$ they have propagated inside the fiber. We do this by switching on the FORT for only a short period of 1-3~ms for loading. This creates an atomic pulse of 1-3~mm length, as can be estimated from the most probable initial velocity $v_0\sim 1$~m/s due to the FORT depth $T_F$ (see Sec.~\ref{sec:axial1}). This is significantly shorter than typical fiber lengths of several centimeters used in experiments. Probing the atoms as a function of time therefore yields information at different positions inside the HCPBGF.

\subsubsection{\label{sec:temperature}Determination of the temperature inside the FORT}
\begin{figure}[t]
\includegraphics[width=8.5cm]{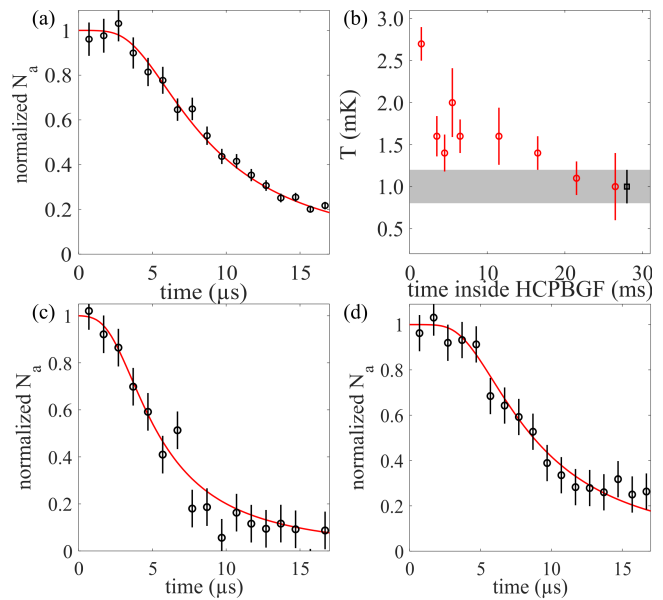}
\caption{\label{fig:tof} (a,c,d) Time-of-flight measurements to determine the ensemble temperature inside the HCPBGF. The normalized atom number ($\circ$) is plotted vs. the time after the FORT has been switched off. The red solid lines show the results of a calculation based on a Gaussian atomic density distribution [see Eq.~\eqref{eqn:tof}]. For maximum loading of the fiber (a) we determine the temperature as $T=1.0(2)~$mK and a radial ensemble $1/e$ half width of $\sigma_a=1.1~\upmu$m for $T_{F}=3.8~$mK. Due to the finite optical pumping time we included a moving average over a $\sim1~\upmu$s period in the calculation. For a pulsed loading of $3$~ms only, the results are shown for times $t_{if}=1.5(15)~$ms (c) and $t_{if}=21.5(15)~$ms (d), respectively. The temperatures obtained are $T=2.7(2)~$mK (c) and $T=1.0(4)~$mK (d). In (b) we plotted the obtained temperatures (\textcolor{red}{$\circ$}) vs. the time $t_{if}$. The temperature from (a) is shown for reference ({\tiny $\square$} and gray-shaded area).}
\end{figure}

We start by determining the temperature of the atomic ensemble inside the HCPBGF which affects the radial density profile according to Eq.~\eqref{eqn:density} and thus inhomogeneous broadening. This can be done by employing a time-of-flight (TOF) measurement, e.g., via a release-and-recapture technique as shown by Bajcsy \textit{et al.} \cite{BHP11}. Here, we simply measure the atom number by TROP after the FORT has been switched off. If the pumping time is much shorter than the time the atoms reach the fiber wall, we verified that this simple method yields the same temperature as the release-and-recapture technique. We assume that the atoms are not detectable once they reached the fiber wall \cite{RZD97,GBR06}. The result of such measurement, taken at $t=1050~$ms, i.e., 5~ms after loading has ended, is shown in Fig.~\ref{fig:tof}(a).
Here, the fiber was loaded to its maximum by keeping the FORT on during the transfer process from the MOT. The atoms have therefore spent $5~$ms $\leq t_{if} \leq$ $35$~ms inside the fiber at the time of the measurement (see Fig.~\ref{fig:Natom_time}). Thus, and because we do not observe an increase of the temperature for $t>1050~$ms, the measured temperature for these conditions ought represent the steady-state temperature. In order to extract the temperature from this data we proceed as follows: Using the measured FORT power, we calculate the FORT potential and radial density according to Eq.~\eqref{eqn:UFORT} and \eqref{eqn:density}. In good approximation, $n(r)$ can be written as a Gaussian of $1/e$-width $\sigma_a$ \cite{SPW19}. After release from the FORT the density evolves as
\begin{equation}
n(r,t) = \frac{N_0}{\pi \sigma_t^2} \exp \left(-\frac{r^2}{\sigma_t^2}\right),
\label{eqn:tof}
\end{equation}
where $\sigma_t^2 = \sigma_a^2 + v_{th}^2t^2$ and $v_{th}=\sqrt{2k_BT/m}$. The dependence of $n$ on the axial coordinate $z$ is irrelevant for this measurement and can therefore be neglected. Integration of Eq.~\eqref{eqn:tof} over $0\leq r \leq r_f$, where $r_f$ is the fiber core radius and normalization leads to the solid red curve in Fig.~\ref{fig:tof}(a). We obtain a temperature of $T=1.0(2)~$mK for a FORT depth of $T_{F}=3.8~$mK. The ratio $T/T_F$ is larger than reported in \cite{BHP11} but smaller than in \cite{YB19} for a similar fiber. We note that the temperature is the only free parameter in our model after the FORT potential has been characterized as shown in App.\ref{sec:characterization}.

Next, we performed TOF measurements with a variable delay after the atoms are loaded for a short period of $3$~ms only. This enables measuring the temperature as a function of time $t_{if}$ the atoms have spent inside the fiber. In Fig.~\ref{fig:tof}(d) we show the results for $t_{if}\sim 21.5$~ms. Comparison to the simulation yields a temperature of $T=1.0(4)~$mK. The temperature agrees well with the one obtained for maximum loading, where the atoms have spent $>5~$ms inside the fiber. If, however, we measure the temperature at the earliest possible time $t_{if}\sim 1.5$~ms after the loading is finished, we obtain a temperature of $T=2.7(2)~$mK. Figure~\ref{fig:tof}(b) depicts the results for the temperatures at different times $t_{if}$ (\textcolor{red}{$\circ$}). The temperature is initially much larger than the steady-state temperature ({\tiny $\square$}) from Fig.~\ref{fig:tof}(a), but converges to this value with time spent inside the fiber. 

We believe this decrease of temperature during the course of about 20-30 ms to be the result of free evaporative cooling inside the FORT, as for instance observed in \cite{BSC01}. As the atoms enter the FORT of finite depth $T_F$, they are accelerated and heated to a maximum of $T \sim T_F$. The trap is therefore overheated and a significant fraction of atoms has a potential energy close to zero. Thus, any heating process present leads to immediate loss of high-energy atoms from the trap and therefore cooling. The rate of evaporation due to elastic collisions can be estimated \cite{KV96} as $\Gamma_{ev}\sim 2$~s$^{-1}$ for our estimated in-fiber atomic density of $10^{12}~$cm$^{-3}$ and trap depth $T_F$. This is too small to explain the observed decrease in temperature. However, as discussed in Sec.~\ref{sec:evolution}, losses from the FORT are currently dominated by heating due to an ellipticity of the FORT polarization \cite{CKC99} leading to a loss rate of 40~s$^{-1}$. This rate matches quite well to the timescale of the here observed decrease of temperature. Thus, experiments sensitive to the ensemble temperature should be carried out with a sufficiently large delay after loading of the fiber is finished to allow for thermalization.

We note that the observation of evaporation on such a short timescale briefly after loading the FORT is facilitated by the ability to selectively probe the atoms inside the HCPBGF. In free-space FORTs, probing is restricted to timescales where the untrapped atoms have fallen out of the interaction volume \cite{BSC01}.

\subsubsection{\label{sec:axial1}Axially resolved measurements: velocity and acceleration}

In deep, focused-beam FORTs, acceleration of the atoms due to radiation pressure from the FORT or other laser beams can be typically neglected as it is dominated by the steep gradient of the FORT potential. In fiber-based FORTs, however, the atoms inside the fiber are basically untrapped along the fiber axis and even modest photon scattering rates can lead to a significant acceleration, affecting the propagation $z(t)$ inside the fiber. Moreover, as the atomic transitions inside the HCPBGF saturate already at a power in the range of a few nanowatts, any not properly blocked beam (e.g., the $0^{\textrm{th}}$ order beams of the AOMs) can have a detrimental effect on the total scattering rate. Thus, a careful study of the atomic motion is advisable, showing, e.g., whether or not the loading efficiency is affected by propagation effects.
To this end, we recorded EIT spectra with anti-collinear probe and control beams for pulsed loading of the fiber. Here, the spectra are maximally susceptible to the atomic motion along the fiber axis as the two-photon resonance is Doppler-shifted by $\sim 2k_{p,c} v_a$. This allows to determine the average axial velocity $v_a(t)=v_0 + a_{eff} t$ as a function of time, and the effective acceleration $a_{eff}$. In the case of a cold atomic cloud held at rest above the fiber tip, the most probable initial velocity after entering the fiber can be estimated from the trap depth $T_F$ as $v_0=\sqrt{2k_B T_F/m}$. The position inside the HCPBGF is then simply given by $z(t) = v_0 t + \frac{1}{2} a_{eff} t^2$. 

\begin{figure}[t]
\includegraphics[width=8.5cm]{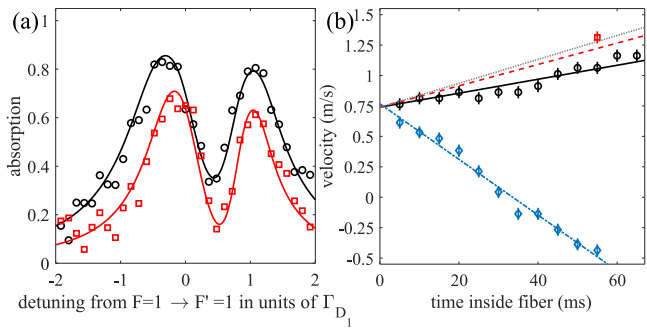}
\caption{\label{fig:velocities} (a) EIT spectra with anti-collinear probe and control fields taken at $t_{if}=35~$ms ($\circ$) and $t_{if}=65~$ms (\textcolor{red}{{\tiny $\square$}}) after the atoms entered the HCPBGF. The solid lines show the results from a numerical simulation with  $d_{opt}=(2.2; 1.3)$ (black; red), $\Omega_c=1.2(1)\Gamma_{D_1}$, $v_a = (0.86; 1.16)$~m/s (black; red), $\gamma_{21}=(0.16;0.07)\Gamma_{D_1}$ (black; red), $T=1~$mK. (b) Measured velocities of the atoms (positive when parallel to the direction of gravity) vs. the time $t_{if}$ they have spent inside the HCPBGF. The estimated uncertainty for determining the velocities is $\pm0.05$~m/s. For regular loading conditions ($\circ$), we obtain $v_0=0.70(6)~$m/s and $a_{eff}=6.8(15)~$m/s$^2$ from a linear least-squares fit.  When the depumper is turned on for only $200~\upmu$s just before the measurement, the effective acceleration is larger (\textcolor{red}{{\tiny $\square$}}). The black solid (red dashed) line shows the result of the rate equation model (see App.~\ref{sec:appendix_RE}) when the depumper is on (off) during loading. Here, we used the measured depumper power, $\Gamma_F^{(R)}=410~$s$^{-1}$, and $v_0=0.74~$m/s which are both within the experimentally determined uncertainties. If the acceleration were solely determined by gravity, the gray dotted line would show the corresponding dependence. When the notch filter used to suppress resonant scattering by the FORT is removed and the depumper is continuously on during the loading process (\textcolor{blue}{$\diamond$}), the atoms are effectively accelerated upwards. The blue dashed-dotted line shows the least-squares linear fit yielding $v_0 = 0.77(8)$~m/s and $a_{eff} = -23(2)~$m/s$^2$.}
\end{figure}

For an atomic cloud entering the HCPBGF at a downward velocity $v_0>0$ and then experiencing solely a downward acceleration in the earth's gravitational field we expect to observe EIT spectra where the two-photon resonance, i.e., the EIT dip, is increasingly shifted to positive detunings with time $t_{if}$ inside the fiber for a downward propagating probe beam.
This is confirmed by EIT spectra taken at $t_{if}=35~$ms ($\circ$) and $t_{if}=65~$ms (\textcolor{red}{{\tiny $\square$}}) as shown in Fig.~\ref{fig:velocities}(a). The solid lines show calculated spectra based on the theory in \cite{BSH16}.
We then extracted the velocity as a function of $t_{if}$ from a series of measurements which is shown in Fig.~\ref{fig:velocities}(b) ($\circ$). A linear fit to the data yields $v_0 = 0.70(6)~$m/s which agrees reasonably well with the expected value due to a FORT depth of $T_{F}=3.8~$mK for this measurement. However, the acceleration $a_{eff}=6.8(15)$m/s$^2$ is smaller than expected from gravity only (gray dotted line). Using this information, we can now calculate the time-dependent position $z(t)$ of the atoms and, e.g., reconstruct for the first time the density evolution inside such a fiber (see Sec.~\ref{sec:evolution}).

As an interesting feature, the measured acceleration inside the fiber is different from the case of free-falling atoms. Radiation pressure from the FORT and depumper beams, which are both aligned anti-parallel to gravity and are typically on during the loading and guiding period, are most likely the source of a reduced acceleration. We therefore recorded spectra when the depumper was switched on for only $200~\upmu$s just before the measurement to avoid any influence on the atomic motion. The extracted velocity is shown by the red square in Fig.~\ref{fig:velocities}(b). It is consistent with the one expected for a purely gravitational acceleration within the error bars.
To further investigate the acceleration, we employed a population rate equation model (see App.~\ref{sec:appendix_OP}), which then allows for calculation of the light pressure force via the photon momentum, excited state scattering rate and population \cite{MS99b}. This model allows for consistently reproducing the optical pumping by the FORT and depumper discussed in App.~\ref{sec:appendix_FORT} as well as the velocities shown in Fig.~\ref{fig:velocities}(b) by the black solid line (depumper always on) and red dashed line (depumper pulsed for $200~\upmu$s). We suspect that the discrepancy between the measured and simulated velocity without the depumper (\textcolor{red}{{\tiny $\square$}}, red dashed line) is due to a slightly higher initial velocity when the depumper is not present, as it will be less-detuned from the cycling transition $F=2 \rightarrow F'=3$ just outside of the HCPBGF where the light-shift by the FORT is smaller and therefore slow the atoms down when it is present. We thus conclude that the effective acceleration inside the fiber is mainly determined by gravity and radiation pressure from the FORT and depumper. Other sources--if present--play only a minor role.

\textit{Notes:} (i) To further illustrate the importance of photon scattering on the propagation dynamics we show in Fig.~\ref{fig:velocities}(b) by blue diamonds the results when a notch filter, used to suppress resonant scattering by the FORT, is removed. This leads to a reversal of the atoms' propagation direction and prohibits a proper filling of the fiber (see Sec.~\ref{sec:evolution}). (ii) Although this EIT configuration with anti-collinear probe/control beams allows for determination of the thermal velocity distribution along the beams' direction, we refrain from a detailed analysis. As we will show in the next section, determination of the ground state dephasing rate shows some irregularities which also affect the precision of determining the temperature. Nonetheless, analysis of spectra such as in Fig.~\ref{fig:velocities}(a) show that the temperature along the fiber axis is in the same range of 1~mK as determined in Sec.~\ref{sec:temperature} for the radial direction.

\subsubsection{\label{sec:axial2}Axially resolved measurements: dephasing}
Information on the ground state dephasing rate $\gamma_{21}$ is vital for experiments requiring long coherence times \cite{XLC19,LXH20,LIW20}. This rate is, e.g., affected by transit-time-broadening, two-photon linewidth of the probe/control beams, and inhomogeneous broadening due to the similar transverse dimensions of the control and the atomic density distribution \cite{BSH16}. In addition, it could depend on the spatial position due to a possible non-zero magnetic field (gradient) which changes across the fiber length of $14~$cm, leading to a spatially varying ground state splitting. EIT spectra taken with collinear beams are minimally sensitive to atomic motion. Thus, spectra taken at different times $t_{if}$ potentially allow for spatially-resolved determination of the effective ground state dephasing rate $\gamma_{21}$.

\begin{figure}[t]
\includegraphics[width=8.5cm]{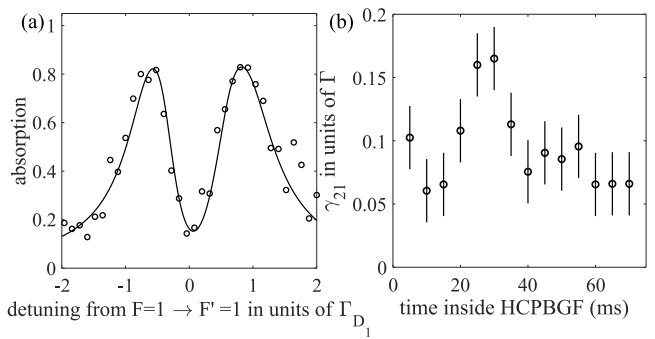}
\caption{\label{fig:decoherence} (a) EIT spectra with collinear probe and control fields taken at $t_{if}=15(1)~$ms ($\circ$) after the atoms entered the HCPBGF. The solid line shows the result of a numerical simulation with $d_{opt}=2.0(1)$, $\Omega_c=1.4(1)\Gamma_{D_1}$, $T=1~$mK, $v_0 = 1.0$~m/s, $\gamma_{21}=0.07(2)\Gamma_{D_1}$. (b) Averaged dephasing rates $\gamma_{21}$ vs. $t_{if}$ as extracted from multiple measurements such as in (a). The error bars represent an estimated uncertainty of $0.02\Gamma_{D_1}$.}
\end{figure}

An EIT spectrum for collinear probe/control beams taken at $t_{if}=15(1)$ is shown in Fig.~\ref{fig:decoherence}(a). As before, we adjusted the parameters of a numerical simulation (solid line) until we reached best agreement with the experimental data ($\circ$) \cite{BSH16}.
Such spectra allow us to extract $\gamma_{21}$ as a function of $t_{if}$ (and therefore position) inside the HCPBGF as shown in Fig.~\ref{fig:decoherence}(b). The lower limit that can be expected for our experiment is given by transit-time broadening leading to $\gamma_{ttb}\sim v_{th}/d_c = 0.013\Gamma_{D_1}$. Obviously, the measured dephasing rates are significantly larger than this lower limit, as already observed in \cite{PWN20}. Additionally, for early as well as later times $t_{if}$, $\gamma_{21}=0.08(2)\Gamma_{D_1}$ is roughly the same. For $20$~ms $\leq t_{if} \leq 35$~ms, however, the dephasing rate increases up to $\gamma_{21}=0.16(2)\Gamma_{D_1}$. We confirmed this behavior during several measurement runs on different days and the data shown here are averages of these multiple measurements. This only temporary increase of the dephasing rate is again quite surprising as we would have expected a steady increase for a spatially varying magnetic field (gradient). This issue is yet undetermined, but again shows the importance of a detailed analysis. 

The data and simulations presented above show that it is indeed possible to gain spatially-resolved information along the fiber axis on the radial distribution, axial position, and ground state dephasing rate for an atomic ensemble loaded into a HCPBGF. This information can for instance be used to understand and optimize the loading of atoms into a fiber as we will show in the following section. We note that for these measurements to succeed with an acceptable signal-to-noise ratio, we required to load around 5000-6000 atoms for a pulsed loading of 2-3~ms duration. 

\subsection{\label{sec:atom_number}Determination of the limits to the loading efficiency}

In the following, we investigate the loading process of the HCPBGF as to determine the current limits to the loading efficiency and prospects for further enhancement. Contrary to other fiber loading techniques \cite{BHP11,HPB18,LWK18}, this has not yet been reported using a dark funnel guide \cite{BHP14}.
Specifically, we investigate the transfer from the MOT to the HCPBGF, the maximum number of atoms $N_{a}$ loaded into the HCPBGF, and their spatio-temporal evolution.

\subsubsection{\label{sec:MOT-HCPBGF_transfer} MOT to HCPBGF atom transfer}

\begin{figure}[t]
\includegraphics[width=7.0cm]{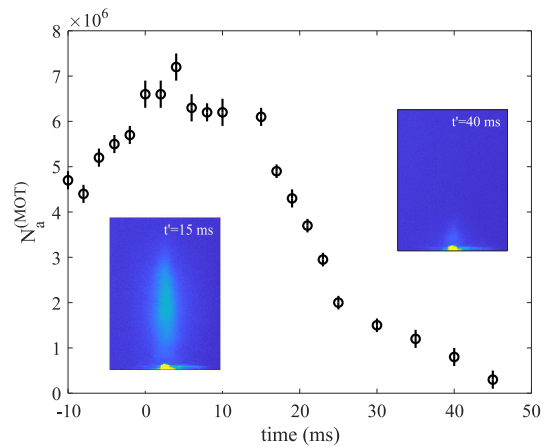}
\caption{\label{fig:Natom_above_HCPBGF} Measured number of atoms $N_a^{(MOT)}$ above the HCPBGF vs. time $t'$ after start of the transfer from the MOT to the HCPBGF. The atom number above the HCPBGF is measured by fluorescence imaging. The error bars correspond to the standard deviation of 20 successive measurements. The MOT repumper is switched off near $t'=0~$ms. The atom cloud reaches the HCPBGF tip around $t'=15~$ms. The MOT quadrupole field is switched off at $t'=45~$ms. The insets show false color images of the atom cloud above the HCPBGF tip (yellow structure at lower end) at two different timings.
}
\end{figure}

In contrast to experiments loading laser-cooled atoms into Kagome-structured hollow fibers of low numerical aperture, we first have to transfer the atoms from the MOT near the HCPBGF tip before the FORT potential is deep enough to guide atoms into the fiber. We do this by employing a DFR to maximize the atomic density above the HCPBGF (see Sec.~\ref{sec:experimental_setup}). We therefore first investigate the transfer of atoms from the MOT (centered $\sim5.5$~mm above the fiber tip) down to the HCPBGF by measuring the number of atoms trapped above the fiber via fluorescence detection \cite{TEC95} with a calibrated CCD camera for different times during the transfer period $990$~ms $\leq t \leq 1050~$ms (see Fig.~\ref{fig:Natom_above_HCPBGF}).
As long as the MOT repumper is on ($t'\lesssim 0~$ms) the number of atoms above the fiber increases. After the MOT repumper is off, leaving only the DFR on, the atom cloud transfer to the HCPBGF occurs without significant losses as long as the cloud has not reached the fiber tip ($0~$ms $\leq t' \leq 15~$ms). Once the lower end of the cloud has reached the HCPBGF tip at $t' \sim 15~$ms, however, the number of atoms decreases quickly due to the losses in the regions where the HCPBGF mount blocks the trapping light. At $t'=45~$ms, where the MOT quadrupole field confining the atoms is switched off, basically no atoms are left above the HCPBGF. These results show that the magnetic offset field settings, which determine the position where the cloud is held during loading, have to be carefully optimized, or losses from the dark funnel guide will become relevant. On the other hand, these magnetic field settings can also be used to quickly change the number of atoms loaded into the HCPBGF while keeping all other parameters the same.
For this peak atom number of $N_a^{(MOT)}=7(1)\times 10^6$ in the MOT we detected $N_a=1.5(1)\times 10^5$ inside the fiber. The loading efficiency from MOT to HCPBGF-based FORT is therefore roughly $\eta=2.1(3)~\%$. The uncertainty of $\eta$ is based on the shot-to-shot fluctuations and does not account for an incorrect calibration of the fluorescence imaging setup. Therefore we expect the uncertainty to be actually larger than $0.3~\%$. The efficiency is nonetheless roughly in agreement with our previously estimated loading efficiency \cite{BHP14} and of similar magnitude as in \cite{HPB18}. 

\subsubsection{\label{sec:FORT_power}Dependence of the loading efficiency on the FORT power}
Given that the MOT to HCPBGF loading efficiency is only a few percent, we might expect that the loading efficiency increases for a deeper FORT potential when the ensemble temperature above the HCPBGF is kept constant. However, we found that this assumption is only partially correct.

\begin{figure}[t]
\includegraphics[width=8.5cm]{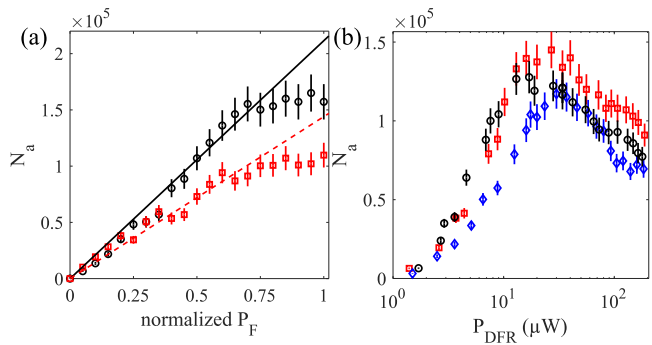}
\caption{\label{fig:Natom_PFORT_DFR} (a) Measured $N_{a}$ vs. relative FORT power $P_{F}/P_{F}^{max}$ with ($\circ$) and without (\textcolor{red}{{\tiny $\square$}}) sub-Doppler cooling above the HCPBGF. The solid lines are linear fits to the corresponding experimental data for $P_{F}/P_{F}^{max} \leq 0.7$. (b) Number of atoms $N_a$ loaded into the HCPBGF vs. the power $P_{DFR}$ of the DFR (of a single beam) for different depumper powers. The error bars in the vertical direction correspond to an estimated $8\%$ uncertainty. The depumper peak Rabi frequency inside the HCPBGF was around $1.4\Gamma_{D_2}$ ($\circ$), $2.9\Gamma_{D_2}$ (\textcolor{red}{{\tiny $\square$}}), and $7.6\Gamma_{D_2}$ (\textcolor{blue}{$\diamond$}), respectively.
}
\end{figure}

In Fig.~\ref{fig:Natom_PFORT_DFR}(a) we measured $N_{a}$ as a function of FORT power $P_{F}$ ($\propto T_F$) during the loading phase. Here we show the results for the case where the trapping detuning is kept constant at $-2.5\Gamma_{D_2}$ (\textcolor{red}{{\tiny $\square$}}) and when the detuning is ramped down to $-5\Gamma_{D_2}$ during the loading process ($\circ$) to optimize sub-Doppler cooling above the HCPBGF.
In both cases the atom number initially increases nearly linearly with the FORT power as expected. For $P_{F}/P_{F}^{max} > 0.7$, however, the atom number saturates around $N_{a}\sim 1.6\times 10^5$ in the case where sub-Doppler cooling is performed. The solid lines are linear fits to the corresponding experimental data for $P_{F}/P_{F}^{max} \leq 0.7$ and are extrapolated until $P_{F}/P_{F}^{max} = 1$. Without sub-Doppler cooling, the linear region extends further, but also here $N_{a}$ saturates near $N_{a}\sim 1.0\times 10^5$. 
We interpret these results as follows: As the maximum atom number inside the HCPBGF seems to saturate at different values for the two conditions, and this maximum value can change from day to day (going sometimes up to $N_{a}\geq 2\times 10^5$), we believe that this limit is due to the density and temperature of the atoms above the fiber. Loading more atoms inside the MOT, better compression and/or cooling during the transfer should therefore lead to a better loading efficiency and is only a technical issue. We note that a similar saturation behavior has also been observed for a different loading technique using a Kagome fiber \cite{HPB18}.

\subsubsection{\label{sec:repumper_power}Dependence of the loading efficiency on the repumper power and configuration}

The DFR \cite{BHP14} is at the heart of our loading method as it guides the atoms and enhances their density near the fiber tip. This distinguishes our loading technique from others and therefore deserves a closer look. 
 
First, we determine how much the DFR affects the loading efficiency. To this end, we compare the loading  using a DFR to the case of a free-falling cloud as well as simple cooling and compression during the transfer. For a free-falling cloud, without any additional compression, cooling  or guiding until the fiber tip, we were not able to detect any atoms inside the HCPBGF, in contrast to \cite{BHP11}. 
When we keep the MOT repumper on during the HCPBGF loading phase and optimize the MOT repumper power to $180~\upmu$W for a maximum $N_{a}$ while the DFR is still blocked, we loaded $N_a=6.6(7)\times10^4$ atoms. This corresponds to a compression and cooling of the cloud during the transfer as it is shifted towards the fiber. Density-limiting collisions are here not suppressed. If, however, we guide the atoms in the dark funnel by switching off the MOT repumper at the beginning of the HCPBGF loading phase ($t=1000~$ms) while keeping the DFR on at a power of $25(1)~\upmu$W per beam, $N_{a}$ increases by a factor of $\sim 2.7$.
This demonstrates the effectiveness of our guiding method based on the DFR. It also nicely matches the results by Kuppens \textit{et al.} who studied the loading of a free-space dipole trap and reported a factor of two improvement in the loading efficiency when using a shadow in the repumper \cite{KCM00}. 

Next, we consider the dependence of $N_{a}$ on the DFR and depumper power. The latter one is relevant as it serves to confine the atoms in $F=1$. Inside the HCPBGF the depumper compensates scattering by the FORT whereas outside of the fiber it compensates scattering by repumper stray light \cite{APE94}. The results are shown in Fig.~\ref{fig:Natom_PFORT_DFR}(b) for three different depumper powers. 
For all datasets $N_{a}$ first increases with DFR power $P_{DFR}$, reaches a maximum and then decreases again. We achieve a maximum $N_{a}$ for $P_{DFR}\sim 25~\upmu$W and a depumper Rabi frequency of $2.9\Gamma_{D_2}$  (\textcolor{red}{{\tiny $\square$}}). The position of the maximum shifts to larger $P_{DFR}$ when the depumper gets stronger. However, the maximum $N_a$ for the strongest depumper (\textcolor{blue}{$\diamond$}) is lower than for optimum conditions (\textcolor{red}{{\tiny $\square$}}). 
An exact interpretation of this behavior is difficult, as the depumper beam serves to pump the atoms both inside and outside of the HCPBGF into $F=1$ where the intensities are very different due to the diverging beams above the fiber. Also, the loading process temporally coincides with propagation of atoms already inside the fiber and thus a temporal modulation of the depumper power for the spatially separated regions is not possible. 
An order-of-magnitude estimation however shows that even for the weakest depumper in Fig.~\ref{fig:Natom_PFORT_DFR}(b) its Rabi frequency is at least comparable to stray light by the DFR over the whole MOT-HCPBGF distance. Therefore, optical pumping into $F=1$ is in principle efficient due to the long pumping times. For regions close to the fiber tip, where the depumper is most crucial, we therefore expect the depumper to effectively confine the population in the ground state $F=1$. Near the fiber tip, however, light-shifts induced by the depumper itself might become relevant that could explain the decrease of the maximum $N_a$ for a stronger depumper (\textcolor{blue}{$\diamond$}).
A more detailed modeling of the loading would be required for a detailed understanding, which is, however, beyond the scope of this work.
Nonetheless, by carefully adjusting the DFR and depumper powers we can achieve optimum loading conditions into the HCPBGF.

\subsubsection{\label{sec:evolution}Temporal evolution of the loading process}

Understanding the dynamics of the loading process is crucial to optimizing the loading efficiency from the MOT into the HCPBGF. As the atoms enter the deep FORT inside the fiber their density increases and density-dependent losses become relevant. The maximum $N_a$ that can be loaded therefore depends on the relative loading and loss rates.

\begin{figure}[b]
\includegraphics[width=8.5cm]{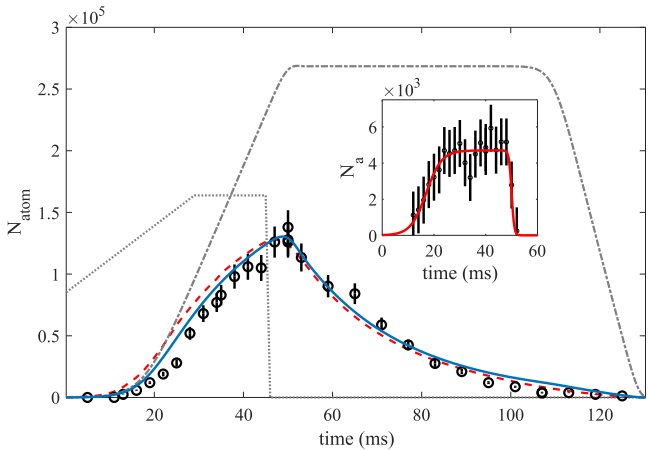}
\caption{\label{fig:Natom_time} Measured $N_{a}$ inside the HCPBGF ($\circ$) vs. time $t'$ after the start of the transfer from the MOT.
The MOT quadrupole field strength is schematically shown by the gray dotted line. The error bars correspond to a $10\%$ measurement uncertainty of $N_a$. Numerical results of the evolution based on Eqns.~\eqref{eqn:con_eq1}-\eqref{eqn:con_eq2} for $a_{eff}=g=+9.81~$m/s$^2$, $\Gamma_L=\beta_L=0$ are shown by the gray dashed-dotted line and for $a_{eff}=+6.8~$m/s$^2$, $v_0=0.8$~m/s, $\Gamma_L=40(5)~$s$^{-1}$ and $\beta_L=1(1)\times10^{-11}~$cm$^3$ s$^{-1}$ by the blue solid line. The loading rate $R(t)$ is taken from the data shown in the inset with an amplitude scaling factor of $S_f=2$ to account for losses during the initial $5~$ms inside the fiber. The red dashed line shows the results when an acceleration of $a_{eff}=-23~$m/s$^2$ inside the HCPBGF is assumed (see text). Inset: $N_{a}$ vs. $t'$ for a loading time of 1~ms. The error bars correspond to an uncertainty in the measurement of 1300 atoms. The red solid line shows the idealized evolution of $N_a$ used for the simulations.}
\end{figure}

To study the loading process in time we modified the sequence described above by adjusting the switch-off time of the quadrupole field and the timings of all laser beams corresponding to the time of each measurement shown in Fig.~\ref{fig:Natom_time} ($\circ$). Here, a time of $t'=0$~ms corresponds to $t=1000$~ms after the MOT loading sequence has started, i.e., loading of the FORT begins. 
At $t' \sim 15$~ms the number of atoms inside the HCPBGF rises rapidly with a rate of roughly $4000$~ms$^{-1}$. According to images taken with a CCD camera (see Fig.~\ref{fig:Natom_above_HCPBGF} inset) this time coincides with the time the lower part of the atom cloud reaches the fiber tip. During the next $\sim20$~ms the rate decreases while $N_{a}$ reaches its maximum value at $t'\sim 50~$ms. This agrees well with the data in Fig.~\ref{fig:Natom_above_HCPBGF} showing the evolution of atoms above the fiber. The maximum $N_{a}$ is reached around 5~ms after the quadrupole field has been switched off. For times $t'>50~$ms $N_{a}$ decreases during the course of around 80~ms. 

To understand this behavior we first need to determine the HCPBGF loading rate. This rate obviously depends on the evolution of the atomic density above the fiber where the FORT potential is deep enough to guide atoms into the HCPBGF. We estimate the region from which atoms are guided into the HCPBGF to be in the range of $200~\upmu$m (axially) and $50~\upmu$m (radially), respectively. As these values are much smaller than the (initial) dimensions of the atomic cloud above the HCPBGF $(l_z,l_r)\sim(3000~\upmu$m$ \times 300~\upmu$m) the measured atom number inside the HCPBGF is proportional to the local density $\rho_{tip}$ of the atom cloud at the fiber tip. 
In our case, the axial velocity of the cloud near the fiber tip can be estimated as $\sim 0.33~$m/s. Thus, by loading the HCPBGF for only a period of 1-3~ms (by pulsing the FORT, see above) the atom cloud moves much less than its length. Measuring $N_{a}$ as a function of time $t'$ therefore yields a value proportional to the local atomic density $\rho_{tip}(t')$ above the HCPBGF.
The result of such measurement is shown in the inset of Fig.~\ref{fig:Natom_time} ($\circ$). 
$N_a(t')$ increases from around 10~ms to 25~ms as the cloud approaches the fiber. Then, the atom number stays roughly constant (neglecting the data point at $t'=34~$ms) for the next $\sim 25$~ms as the cloud is held near the fiber tip. After the MOT quadrupole field is switched off at $45~$ms $N_a(t')$ decreases quickly for $t'>48~$ms. 
We thus estimate the peak loading rate as $\sim 4700~$ms$^{-1}$ which is in reasonable agreement with the initial loading rate obtained from Fig.~\ref{fig:Natom_time}.

The data shown in the inset of Fig.~\ref{fig:Natom_time} can now be used to simulate how $N_{a}$ is expected to evolve in time. To this end we used an idealized atom number/density evolution shown by the red solid line.
In regular focused-beam FORTs of large $N_a$ the evolution of trapped atoms could be modeled according to 
\begin{equation}
	\frac{dN_{a}}{dt} = R'(t) -\Gamma_L' N_{a} - \beta_L' N_{a}^2,\label{eqn:losses}
\end{equation}
where $R'(t)$ is the time-dependent loading rate, $\Gamma_L'$ describes losses due to collisions with background gas as well as heating and $\beta_L'$ describes two-body collisions, respectively \cite{KCM00}.
In the case of our extended fiber-based FORT, however, Eq.~\eqref{eqn:losses} will not suffice. In addition, we have to take propagation of the atoms through the fiber as well as their acceleration (see Sec.~\ref{sec:axial1}) into account. We therefore employ the following continuity equation with loss terms and consider a changing atomic velocity due to an effective acceleration $a_{eff}$ as measured in Sec.~\ref{sec:axial1}
\begin{eqnarray}
	\frac{\partial}{\partial t}n + \frac{\partial}{\partial z}\left(n\cdot v_a\right) &=& -\Gamma_L n - \beta_L n^2,\label{eqn:con_eq1}\\
	\frac{\partial}{\partial t}v_a &=& a_{eff}.\label{eqn:con_eq2}
\end{eqnarray}
Here, the unprimed rates $\Gamma_L$, $\beta_L$ refer to the loss of density instead of atom number. Using the loading rate $R'(t)$ as shown in the inset of Fig.~\ref{fig:Natom_time} by the red line as initial condition at the fiber input (after conversion into a density rate $r(t)$ using the known FORT potential and atomic temperature), as well as an initial velocity of $v_0=+0.8~$m/s [see Fig.~\ref{fig:velocities}(b)] we can solve Eqns.~\eqref{eqn:con_eq1}-\eqref{eqn:con_eq2} either numerically or analytically (see App.~\ref{sec:appendix_evo}) yielding
\begin{equation}
 n(z,t)={  e^{-\Gamma_L T(z)}\over    1 +{\beta_L r(t-T(z))\over v_0\Gamma_L}( 1 -   e^{-\Gamma_L T(z)})}{r(t-T(z))\over v_0},
\end{equation}
where the time $T(z)$ is defined in the appendix. 
Finally, the density is converted into an atom number $N_a$ by integration over the axial and radial dimensions.
Since the data in the inset of Fig.~\ref{fig:Natom_time} was obtained $5~$ms after the atoms entered the fiber, losses are present during this period that have to be accounted for by a scaling factor $S_f$ in the initial loading rate for the simulations. This mainly affects the peak $N_a$ that can be loaded.

The gray dashed-dotted line in Fig.~\ref{fig:Natom_time} shows the result for ideal conditions with an acceleration of +9.81~m/s$^2$ due to gravity and vanishing loss rates $\Gamma_L,\, \beta_L$. Initially, the agreement with the measured data is reasonable. For longer times, however, a huge discrepancy in terms of maximum $N_a$ and decay is obvious that illustrates the importance of losses and heating \cite{KCM00,CKC99}.
The result which matches our experimental data ($\circ$) best is shown by the blue solid line with $\Gamma_L=40(5)$~s$^{-1}$ and $\beta_L = 1(1)\times 10^{-11}$~cm$^3$ s$^{-1}$, and $S_f=2$. An animation of the corresponding spatio-temporal density evolution can be found \href{https://doi.org/10.6084/m9.figshare.14184872.v1}{here}. As we can see, even for optimum conditions only about $30~\%$ of the fiber are filled with atoms and the axial density is far from homogeneous. The rate $\Gamma_L$ clearly dominates the losses, which is about 14 and 5 times larger than reported by Okaba \textit{et al.} \cite{OTB14} and Hilton \textit{et al.} \cite{HPB18}, respectively, for Kagome-structured fibers. It is also significantly larger than reported by Kuppens \textit{et al.} for a free-space trap, where they found losses being dominated by $\beta_L \sim 4\times 10^{-11}$~cm$^3$ s$^{-1}$ while $\Gamma_L$ was found to be near zero \cite{KCM00}. Due to the large $\Gamma_L$ in our case, we can only determine the upper limit of as $\beta_L < 2\times 10^{-11}$~cm$^3$ s$^{-1}$. This value is roughly in agreement with the lower limits reported in \cite{HPB18} and \cite{KCM00} and could currently be limited by light-assisted collisions due to the depumper.

The loss rate $\Gamma_L$ is determined by collisions of trapped atoms with background gas inside the fiber as well as heating. In both studies using Kagome-structured fibers \cite{HPB18,OTB14}, collisional losses due to residual gases inside the fiber were assumed to dominate $\Gamma_L$ while heating was neglected. Heating by photon scattering of the FORT can be estimated for our trap as $\dot{T} \sim 10~\upmu$K s$^{-1}$ \cite{GWO00} and is therefore negligible on the considered timescale. However, heating induced losses due to a possible elliptical polarization of the FORT, as discussed by Corwin \textit{et al.} \cite{CKC99}, resulting in a hopping between Zeeman levels of different potential depths, can be estimated as around $57(23)$~s$^{-1}$. This fits quite well to the loss rate determined above from Fig.~\ref{fig:Natom_time}. Here we assumed an ellipticity of the FORT inside the HCPBGF of 0.9, a hopping rate of $690(280)$~s$^{-1}$ according to the results in App.~\ref{sec:appendix_FORT}, potentials for states $F=1,\,m_F=+1$ and $F=2,\,m_F=+2$ according to Eq.~\eqref{eqn:UFORT}, and initial and final temperatures of $T_i=1~$mK and $T_f=4.0~$mK, respectively. The ellipticity of the FORT is estimated and based on the fact that though our fiber can maintain linear polarizations quite well at the output for a certain input alignment, inside the fiber the polarization is slightly elliptical due to a circular birefringence component \cite{PWN20}. 

\textit{Notes}: (i) The agreement between numerical and experimental results is quite good despite the simplicity of our model. (ii) The loss rate $\Gamma_L$, can be explained well by heating and the rate $\beta_L$ is of similar magnitude as in other FORTs. It also explains well the time required to reach the steady-state temperature as shown in Fig.~\ref{fig:tof}(b). (iii) Currently, $N_a$ is limited by evaporation from the FORT. In-fiber cooling (during the loading and guiding process) might therefore lead to a larger loading efficiency. (iv) Optical pumping between the hyperfine ground states via a Raman coupling by the broadband FORT (see App.~\ref{sec:appendix_FORT}) significantly affects the acceleration for our loading scheme. Employing a narrowband FORT is therefore crucial when using such system for interferometry \cite{XLC18}. (v) As was shown in Sec.~\ref{sec:axial1}, the atoms experience an acceleration of $a=-23~$m/s inside the fiber and reverse their direction when resonant scattering by the FORT is not suppressed. This was the case in our previous work \cite{PWN20}. Suppression of resonant scattering surprisingly did not affect the loading efficiency. 
We therefore tried to simulate these conditions. However, the theoretical model used to simulate the temporal and spatial density evolution can only be used when the atoms are traveling into a single direction. To model the case of direction reversal we had to employ a cruder model which divides the initial loading rate into slices, each assigned with a distinct velocity that would change in each time step according to the acceleration. The results are shown in Fig.~\ref{fig:Natom_time} by the red dashed line. Interestingly, the results are quite similar to the case of a falling cloud without reversal of direction. This is due to the fact that the atoms are pushed out of the fiber only at late times where losses, dominated by heating and collisions, have already significantly reduced $N_a$ as illustrated in an \href{https://doi.org/10.6084/m9.figshare.14184884.v1}{animation} of the spatio-temporal density evolution. Only up to $15~\%$ of the fiber are filled with atoms under these conditions and the density is temporarily increased significantly.

We conclude that the loading efficiency is currently limited by the number of atoms and the density above the HCPBGF as well as the loss rates inside the fiber. Using a deeper FORT, however, will not lead to a further improvement. Especially the loss rate $\Gamma_L$ seems to be dominated by heating inside the slightly elliptical FORT. As this ellipticity cannot be avoided for our fiber, in-fiber cooling seems to be the most effective way to reduce the losses. According to the simulation, the loading efficiency could be easily increased by a factor beyond 1.8 for loss rates such as reported in \cite{SWM89}.

\section{\label{sec:summary}Summary and Outlook}

We presented thorough experimental investigations, accompanied by theoretical calculations to carefully characterize and optimize the loading of laser-cooled atoms from a MOT into a FORT inside a hollow-core fiber. Although optical access in such a fiber is restricted to the input and output ports, we demonstrated spatially-resolved probing of the atoms inside the fiber with a resolution of a few millimeters using EIT and a time-of-flight technique. 
We observed for the first time a decrease of the atomic ensemble temperature during the course of around 20~ms after entering the fiber, reaching a steady-state value of about $1/4$ of the trap depth. Also, we determined an in-fiber acceleration slightly smaller than expected from gravity which could be explained by radiation pressure. Finally, we determined a peak number density of $\sim 10^{12}~$cm$^{-3}$ and loss rates of 40(5)~s$^{-1}$, dominated by heating, and $1(1)\times10^{-11}~$cm$^3$~s$^{-1}$ due to two-body collisions. Using these results, we were able to reconstruct the spatio-temporal density evolution inside such fiber for the first time. We found good agreement between the experimental data and the simple theoretical models used. 

Although our results for our fiber-guided FORT basically confirm the expectations for free-space FORTs, there are also some significant differences. Most notably, propagation of the atoms inside the extended fiber has to be taken into account, which is not required for standard focused-beam FORTs. Although the loading efficiency is currently not limited by propagation through our fiber, it has to be considered for shorter fibers or lower loss rates. 

Our results allow for optimizing the loading efficiency into such hollow-core fiber and thus have direct implications for applications. First of all, the results on the loss rates show that there is significant room for improvement on our record $d_{opt}=1000$ achieved so far \cite{BHP14}. An even larger $d_{opt}$ would be required for experiments on many-body physics with light \cite{CGM08,NA17}, which seems to be feasible.
Secondly, as the loading efficiency is in part limited by heating and density-dependent losses, in-fiber cooling \cite{LXC20,NEC20} will have to be applied with care. On the one hand, cooling will lead to a reduction of inhomogenous broadening and compensation of heating. On the other hand, the atomic number density will also increase if the FORT potential is kept constant, leading to increased losses. Thus, careful modeling and characterization of the loading will be required for achieving optimum conditions, as we have done in our present work.
Finally, our results on the atomic velocity inside the fiber show that a careful analysis is required when such a platform is used for atom interferometry \cite{XLC18}.

\begin{acknowledgments}
We thank M. Schlosser for helpful discussions on the FORT bandwidth and comments on the manuscript, C.-Y. Hsu and M. Coelle for assistance with time-of-flight measurements, and the group of T. Walther for providing us with a home-made ultra-low noise laser diode driver with high modulation bandwidth. This project has received funding from the European Union’s Horizon 2020 research and innovation programme under the Marie Sklodowska Curie grant agreement No. 765075.
\end{acknowledgments}

\appendix

\section{\label{sec:characterization}Characterization of the FORT}
\label{sec:appendix_FORT}

We here summarize the results of the characterization of the FORT guiding the atoms into the HCPBGF. Precise knowledge of the FORT potential is required for a quantitative analysis of the measurements shown in the main part of the paper, as it affects the atomic temperature, peak number density and its distribution, resonance shifts, scattering rate, and collisions \cite{GWO00}. 

For far-detuned traps, where the FORT detuning is much larger than the hyperfine and fine structure splitting, the following radial potential is applicable for the ground state \cite{GWO00,CKC99}
\begin{eqnarray}
U_{F}(r) = -\frac{\pi c^2 }{2} & & \left[\frac{\Gamma_{D_2}\left(2 + g_F m_F \sqrt{1-\epsilon}\right)}{\omega_{D_2}^3 \left(\omega_{D_2}-\omega\right)}\right. \nonumber \\ 
 & & \left. + \frac{\Gamma_{D_1}\left(1- g_F m_F \sqrt{1-\epsilon}\right)}{\omega_{D_1}^3 \left(\omega_{D_1}-\omega\right)}\right] I(r).
\label{eqn:UFORT}
\end{eqnarray}
Here $\Gamma_{D_{1,2}}$ and $\omega_{D_{1,2}}$ are the excited state decay rates and corresponding transition frequencies, of the D$_{1,2}$ lines, $g_F$ is the Land\'{e} factor, $m_F$ is the magnetic quantum number, $\omega$ is the frequency of the FORT field, and $I(r)$ is its radial intensity profile. The ellipticity $\epsilon$ of the polarization is defined as $\epsilon=(P_{max}-P_{min})/(P_{max}+P_{min})$, where $P_{max,min}$ are the maximum/minimum transmitted powers through a rotated linear polarizer. 

The FORT potential directly determines the atomic level shifts via $\Delta_{ls}(r)=U_{F}(r)/\hbar$, with the reduced Planck constant $\hbar$, and also the thermalized atomic number density distribution \cite{GWO00} (see also Sec.~\ref{sec:temperature})
\begin{equation}
n(r)=n_0\, \exp{\left(-\frac{U_{F}(r)-U_{F}(0)}{k_BT}\right)}, 
\label{eqn:density}
\end{equation}
where $n_0$ is the peak density and $T<T_F$ is the ensemble temperature inside the deep FORT. Therefore, spectroscopy of the atoms while the FORT is on will yield information on the FORT depth, density distribution and temperature \cite{PBH12,HLL20}. We confirmed experimentally via light-shift spectroscopy \cite{HLL20} that Eqns.~\eqref{eqn:UFORT}-\eqref{eqn:density} would predict the correct spectroscopic shifts for the measured FORT power, wavelength and beam width.

An important set of parameters of a cylindrically symmetric FORT are the trap frequencies $\omega_{r,z}$ \cite{GWO00}. Already small disturbances of the trapping potential occurring at frequencies $\omega_{mod}=2\omega_{r,z}/n$ with $n=1,2,...$ 
will lead to parametric heating and losses from the trap \cite{SOT97,FDW98}. This is crucial in our experiment, as we intentionally modulate the FORT on/off for probing the atoms without inhomogeneous broadening induced by the FORT \cite{BHP11,BSH16,PWN20}. 
This modulation frequency should thus be chosen such that parametric heating and loss does not occur. When the FORT is inside a HCPBGF, the axial trap frequency cannot be defined due to the constant potential along the fiber axis leaving only the radial frequency where the potential is approximately harmonic. 
To investigate the FORT trap frequency we followed the method of Friebel \textit{et al.} \cite{FDW98}. We modulated the trapping potential sinusoidally with different modulation frequencies $\omega_{mod}$ and then measured the remaining number of atoms left in the FORT. The experimentally observed resonances are slightly red-shifted, but otherwise agree well with the calculated radial resonance frequencies at $\omega_{mod}=(\omega_r,2\omega_r)$, where $\omega_r=2\pi\times51$~kHz. Interestingly, we observed an additional less-pronounced resonance near $4\omega_r$. This resonance might be due to the occupation of high-lying vibrational levels where the harmonic approximation of the FORT potential is not valid anymore, as known for optical lattices \cite{RJS01,JPR01}.

Photon scattering potentially results in heating and loss as well as unwanted population dynamics. As to avoid hyperfine-changing collisions \cite{SWM89} that could lead to additional losses from the FORT inside the HCPBGF ($h/k_b \cdot 6.835~$GHz$ = 328$~mK $\gg T_{F}$) we keep a depumper beam, tuned near the transition $5 ^2S_{1/2}\,F=2 \rightarrow 5 ^2P_{3/2}\, F'=2$, on whenever the FORT is on. This confines the population in $F=1$ inside and outside of the HCPBGF. Population redistribution between the two hyperfine ground states should therefore only be due to the FORT and depumper.
We thus measured the population $N_a(t)$ in $F=1$ as a function of time after the depumper had been switched off while only the FORT was on. As the FORT couples both ground states off-resonantly to the excited states, optical pumping occurs with an effective relaxation rate $\gamma$, leading to total depolarization of the population according to $N_{a}(t)=N_1+N_2\cdot \exp(-\gamma t)$. A fit to the measured data yields $\gamma = 690(280)$~s$^{-1}$, $N_1=0.34(8), N_2 = 0.66(8)$. 
Next, we  measured the power spectrum of the FORT laser system with an optical spectrum analyzer and calculated the spontaneous FORT scattering rate according to
\begin{equation}
\begin{split}
\Gamma_{F} = \frac{\pi c^2}{2\hbar} \displaystyle\int \left( \frac{2\Gamma_{D_2}^2}{\omega_{D_2}^3 \left( \omega_{D_2}-\omega\right)^2}+ \frac{\Gamma_{D_1}^2}{\omega_{D_1}^3 \left( \omega_{D_1}-\omega\right)^2}\right)\\
I'(\omega) d\omega,\label{eqn:FORT_scat}
\end{split}
\end{equation}
where $I'(\omega)=dI'(\omega)/d\omega$ is the intensity per frequency interval $d\omega$, we assumed linear polarization and that the hyperfine structure is not resolved. We obtained $\Gamma_F\sim 186~$s$^{-1}$, i.e., smaller than $\gamma$.
Care must be taken when comparing the FORT scattering rate $\Gamma_F$ to the population relaxation rate $\gamma$. As was first pointed out by Cline \textit{et al.}, the relaxation rate can be orders of magnitude smaller than the total spontaneous scattering rate, due to an interference between the D$_{1,2}$ lines suppressing Raman scattering \cite{CMM94}. When we consider this interference effect, we obtain a relaxation rate of $\gamma_{rel}=0.07\Gamma_F$ between $F=1,2$, i.e., about fifty times smaller than measured.
To solve this discrepancy, we note that our FORT system is comprised of two independent and unstabilized laser diodes. As a result, the FORT field intensity shows fluctuations, whose spectral components can drive Raman transitions between $F=1,2$ without radiation of spontaneously emitted photons. We will show and discuss the validity of this assumption in Sec.~\ref{sec:axial1}.

\section{Simulation of optical pumping and acceleration by the FORT and depumper beams}
\label{sec:appendix_OP} 

We here present the model used to simulate the data on the axial velocity shown in Fig.~\ref{fig:velocities}(b). 
Both FORT and depumper are basically always switched on during loading of the HCPBGF. Therefore, optical pumping by both fields has to be considered as it affects the population redistribution between the two ground states $F=1,2$ (see App.~\ref{sec:appendix_FORT}) as well as the acceleration $a=F/m$.
Here,
$F = \hbar k \Gamma \rho_{ee}$
 is the force on an atom induced by absorption of photons followed by spontaneous emission at rate $\Gamma$ and $\rho_{ee}$ is the population in the excited state \cite{MS99b}. 

\label{sec:appendix_RE}

In order to calculate the relevant state-resolved populations in our multi-level system, we numerically solved the following population rate equations, where the index $i$ corresponds to the following hyperfine states in ascending energetic order (ground states $F=1,2$, D$_2$ line $F'=1,2,3$). In this model we assume that population is mainly redistributed between $F=1,2$ by Raman scattering at rate $\Gamma_{F}^{(R)}=690(280)~$s$^{-1}$ ($\equiv \gamma$ within the measurement uncertainty, see App.~\ref{sec:appendix_FORT}) while spontaneous scattering of the FORT at rate $\Gamma_{F}=186~$s$^{-1}$ leads predominantly to Rayleigh scattering due to an interference effect between the D$_{1,2}$ lines \cite{CMM94} and therefore does not affect the populations. We thus only need to consider the ground state populations as well as the excited states coupled by the depumper at the D$_2$ line. $\Omega_d(\delta_i)$ are the depumper Rabi frequencies depending on the detuning from levels $F'=1,2,3$. The branching ratios are taken from \cite{Steck2015}. Furthermore, we assumed an unpolarized FORT and depumper, i.e., we neglected the Zeeman substructure and only considered stimulated emission by the near-resonant depumper. The population rate equations then read
\begin{widetext}
\begin{eqnarray}
\dot{N}_1 & = & \Gamma_{D_2} \left(\frac{5}{6} N_3 + \frac{1}{2} N_4 \right) - \frac{5}{8} \Gamma_{F}^{(R)} N_1 + \frac{3}{8} \Gamma_{F}^{(R)} N_2, \label{eq:re_1}\\
\dot{N}_2 & = & \Gamma_{D_2} \left( \frac{1}{6} N_3 + \frac{1}{2} N_4 + 1 N_5 \right)+ \frac{5}{8} \Gamma_{F}^{(R)} N_1 - \frac{3}{8} \Gamma_{F}^{(R)} N_2,\\\nonumber
& &  - \Omega_d(\delta_1)\left(  N_2 - N_3 \right) - \Omega_d(\delta_2) \left( N_2 - N_4 \right) - \Omega_d(\delta_3) \left( N_2 - N_5 \right),\\
\dot{N}_3 & = & -\Gamma_{D_2} N_3 + \Omega_d(\delta_1)\left(  N_2 - N_3 \right),\\
\dot{N}_4 & = & -\Gamma_{D_2} N_4 + \Omega_d(\delta_2) \left( N_2 - N_4 \right),\\
\dot{N}_5 & = & -\Gamma_{D_2} N_5 + \Omega_d(\delta_3) \left( N_2 - N_5 \right).\label{eq:re_12}
\end{eqnarray}
\end{widetext}

\label{sec:appendix_acc}

The total FORT scattering rate $\Gamma_F$ has a direct influence on the change of atomic momentum. The Raman pumping rate $\Gamma_F^{(R)}$ only affects the momentum if the depumper is on, as  the net momentum transfer of $\pm h/c \cdot 6.8~$GHz is zero over many pumping cycles. The rate equations \eqref{eq:re_1}-\eqref{eq:re_12} can be either solved numerically, where the effective acceleration $a_{eff}=g - F/m$ of the atoms is then given by
\begin{equation}
a_{eff} = g - \frac{\hbar k}{m} \left( \Gamma_F + \Gamma_{D_2} \sum_{i=3}^5 N_i\right),
\end{equation}
or analytically by using the stationary solutions. Then, the acceleration can be determined as
\begin{equation}
a_{eff} = g - \frac{\hbar k}{m} \left( \Gamma_F + \frac{5}{8} \Gamma_F^{(R)} p_0\right),
\end{equation}
where the first term is due to downward acceleration by gravity and the other terms due to the upward directed light pressure by the FORT and depumper beams. The parameter $p_0$ is between 1.5 and 2 and depends on the detuning of the depumper from the transition $F=1 \rightarrow F'=2$. For the previously used FORT scattering rates we thus obtain $a_{eff}\sim 5~$m/s$^2$, close to the measured value.

\section{Analytic derivation of the density evolution inside the fiber}
\label{sec:appendix_evo}
We consider the propagation of an ensemble inside a fiber where atoms are moving with positive acceleration into a single direction, i.e., there is no change of direction. The atomic number density $n(z,t)$ inside the fiber at time $t$ in position $z$ is determined by the atomic number density at the beginning $n(z=0,t)$ at the previous time $t-T(z)$, where $T(z)$ is the time to travel from $z=0$ to $z$ with the acceleration $a_{eff}$.
This time is a solution of the equation
\begin{equation}
 z=v_0 T + \frac{1}{2} a_{eff} T^2,
\end{equation}
where $v_0$ is the initial velocity.
The solution of this equation reads
\begin{equation}
 T(z)={z\over v_0}{   2      \over  \sqrt{1+2 a_{eff}z/v_0^2}+ 1 } .
\end{equation}
If the fiber is loaded with the density rate $r(t)$, then the density $n(z=0,t)$ is given by
\begin{equation}
 n(z=0,t)=r(t)/v_0.
\end{equation}
Thus, the density $n(z,t)$ at time $t$ and position $z$ is equal to the density $n(z=0,t-T(z))$ but subject to collisional losses during the time $T(z)$. The differential equation for this density $\nu(\tau)= n(z=0,t-\tau)$ reads
\begin{equation}
 {d \nu(\tau) \over d\tau}=-\Gamma_L \nu(\tau)-\beta_L  \nu^2(\tau).
\end{equation}
with the initial condition $\nu(\tau=0)=r(t-T(z))/v_0$.
The analytical solution of this equation is given by 
\begin{eqnarray}
 \nu(\tau) &= &{\Gamma_L e^{-\Gamma_L \tau} \nu(\tau=0)\over    \Gamma_L +\beta_L\nu(\tau=0)( 1 -   e^{-\Gamma_L \tau})}\\\nonumber
 &=&{  e^{-\Gamma_L \tau}\over    1 +{\beta_L r(t-T)\over v_0\Gamma_L}( 1 -   e^{-\Gamma_L \tau})}{r(t-T)\over v_0}
\end{eqnarray}
and the corresponding solution for $n(z,t)$ is
\begin{equation}
 n(z,t)={  e^{-\Gamma_L T(z)}\over    1 +{\beta_L r(t-T(z))\over v_0\Gamma_L}( 1 -   e^{-\Gamma_L T(z)})}{r(t-T(z))\over v_0}.
\end{equation}
The total atom number $N_a(t)$ is then given by the integral of $n(z,t)$ along the radial and axial directions.


\begin{thebibliography}{61}%
\makeatletter
\providecommand \@ifxundefined [1]{%
 \@ifx{#1\undefined}
}%
\providecommand \@ifnum [1]{%
 \ifnum #1\expandafter \@firstoftwo
 \else \expandafter \@secondoftwo
 \fi
}%
\providecommand \@ifx [1]{%
 \ifx #1\expandafter \@firstoftwo
 \else \expandafter \@secondoftwo
 \fi
}%
\providecommand \natexlab [1]{#1}%
\providecommand \enquote  [1]{``#1''}%
\providecommand \bibnamefont  [1]{#1}%
\providecommand \bibfnamefont [1]{#1}%
\providecommand \citenamefont [1]{#1}%
\providecommand \href@noop [0]{\@secondoftwo}%
\providecommand \href [0]{\begingroup \@sanitize@url \@href}%
\providecommand \@href[1]{\@@startlink{#1}\@@href}%
\providecommand \@@href[1]{\endgroup#1\@@endlink}%
\providecommand \@sanitize@url [0]{\catcode `\\12\catcode `\$12\catcode
  `\&12\catcode `\#12\catcode `\^12\catcode `\_12\catcode `\%12\relax}%
\providecommand \@@startlink[1]{}%
\providecommand \@@endlink[0]{}%
\providecommand \url  [0]{\begingroup\@sanitize@url \@url }%
\providecommand \@url [1]{\endgroup\@href {#1}{\urlprefix }}%
\providecommand \urlprefix  [0]{URL }%
\providecommand \Eprint [0]{\href }%
\providecommand \doibase [0]{http://dx.doi.org/}%
\providecommand \selectlanguage [0]{\@gobble}%
\providecommand \bibinfo  [0]{\@secondoftwo}%
\providecommand \bibfield  [0]{\@secondoftwo}%
\providecommand \translation [1]{[#1]}%
\providecommand \BibitemOpen [0]{}%
\providecommand \bibitemStop [0]{}%
\providecommand \bibitemNoStop [0]{.\EOS\space}%
\providecommand \EOS [0]{\spacefactor3000\relax}%
\providecommand \BibitemShut  [1]{\csname bibitem#1\endcsname}%
\let\auto@bib@innerbib\@empty
\bibitem [{\citenamefont {Ol'Shanii}\ \emph {et~al.}(1993)\citenamefont
  {Ol'Shanii}, \citenamefont {Ovchinnikov},\ and\ \citenamefont
  {Letokhov}}]{OOL93}%
  \BibitemOpen
  \bibfield  {author} {\bibinfo {author} {\bibfnamefont {M.}~\bibnamefont
  {Ol'Shanii}}, \bibinfo {author} {\bibfnamefont {Y.}~\bibnamefont
  {Ovchinnikov}}, \ and\ \bibinfo {author} {\bibfnamefont {V.}~\bibnamefont
  {Letokhov}},\ }\href {\doibase 10.1016/0030-4018(93)90761-S} {\bibfield
  {journal} {\bibinfo  {journal} {Optics Communications}\ }\textbf {\bibinfo
  {volume} {98}},\ \bibinfo {pages} {77} (\bibinfo {year} {1993})}\BibitemShut
  {NoStop}%
\bibitem [{\citenamefont {Debord}\ \emph {et~al.}(2019)\citenamefont {Debord},
  \citenamefont {Amrani}, \citenamefont {Vincetti}, \citenamefont
  {G{\'{e}}r{\^{o}}me},\ and\ \citenamefont {Benabid}}]{DAV19}%
  \BibitemOpen
  \bibfield  {author} {\bibinfo {author} {\bibfnamefont {B.}~\bibnamefont
  {Debord}}, \bibinfo {author} {\bibfnamefont {F.}~\bibnamefont {Amrani}},
  \bibinfo {author} {\bibfnamefont {L.}~\bibnamefont {Vincetti}}, \bibinfo
  {author} {\bibfnamefont {F.}~\bibnamefont {G{\'{e}}r{\^{o}}me}}, \ and\
  \bibinfo {author} {\bibfnamefont {F.}~\bibnamefont {Benabid}},\ }\href
  {\doibase 10.3390/fib7020016} {\bibfield  {journal} {\bibinfo  {journal}
  {Fibers}\ }\textbf {\bibinfo {volume} {7}},\ \bibinfo {pages} {16} (\bibinfo
  {year} {2019})}\BibitemShut {NoStop}%
\bibitem [{\citenamefont {Renn}\ \emph {et~al.}(1995)\citenamefont {Renn},
  \citenamefont {Montgomery}, \citenamefont {Vdovin}, \citenamefont {Anderson},
  \citenamefont {Wieman},\ and\ \citenamefont {Cornell}}]{RMV95}%
  \BibitemOpen
  \bibfield  {author} {\bibinfo {author} {\bibfnamefont {M.~J.}\ \bibnamefont
  {Renn}}, \bibinfo {author} {\bibfnamefont {D.}~\bibnamefont {Montgomery}},
  \bibinfo {author} {\bibfnamefont {O.}~\bibnamefont {Vdovin}}, \bibinfo
  {author} {\bibfnamefont {D.~Z.}\ \bibnamefont {Anderson}}, \bibinfo {author}
  {\bibfnamefont {C.~E.}\ \bibnamefont {Wieman}}, \ and\ \bibinfo {author}
  {\bibfnamefont {E.~A.}\ \bibnamefont {Cornell}},\ }\href {\doibase
  10.1103/PhysRevLett.75.3253} {\bibfield  {journal} {\bibinfo  {journal}
  {Physical Review Letters}\ }\textbf {\bibinfo {volume} {75}},\ \bibinfo
  {pages} {3253} (\bibinfo {year} {1995})}\BibitemShut {NoStop}%
\bibitem [{\citenamefont {Renn}\ \emph {et~al.}(1996)\citenamefont {Renn},
  \citenamefont {Donley}, \citenamefont {Cornell}, \citenamefont {Wieman},\
  and\ \citenamefont {Anderson}}]{RDC96}%
  \BibitemOpen
  \bibfield  {author} {\bibinfo {author} {\bibfnamefont {M.~J.}\ \bibnamefont
  {Renn}}, \bibinfo {author} {\bibfnamefont {E.~A.}\ \bibnamefont {Donley}},
  \bibinfo {author} {\bibfnamefont {E.~A.}\ \bibnamefont {Cornell}}, \bibinfo
  {author} {\bibfnamefont {C.~E.}\ \bibnamefont {Wieman}}, \ and\ \bibinfo
  {author} {\bibfnamefont {D.~Z.}\ \bibnamefont {Anderson}},\ }\href {\doibase
  10.1103/PhysRevA.53.R648} {\bibfield  {journal} {\bibinfo  {journal}
  {Physical Review A}\ }\textbf {\bibinfo {volume} {53}},\ \bibinfo {pages}
  {R648} (\bibinfo {year} {1996})}\BibitemShut {NoStop}%
\bibitem [{\citenamefont {Ito}\ \emph {et~al.}(1996)\citenamefont {Ito},
  \citenamefont {Nakata}, \citenamefont {Sakaki}, \citenamefont {Ohtsu},
  \citenamefont {Lee},\ and\ \citenamefont {Jhe}}]{INS96}%
  \BibitemOpen
  \bibfield  {author} {\bibinfo {author} {\bibfnamefont {H.}~\bibnamefont
  {Ito}}, \bibinfo {author} {\bibfnamefont {T.}~\bibnamefont {Nakata}},
  \bibinfo {author} {\bibfnamefont {K.}~\bibnamefont {Sakaki}}, \bibinfo
  {author} {\bibfnamefont {M.}~\bibnamefont {Ohtsu}}, \bibinfo {author}
  {\bibfnamefont {K.~I.}\ \bibnamefont {Lee}}, \ and\ \bibinfo {author}
  {\bibfnamefont {W.}~\bibnamefont {Jhe}},\ }\href {\doibase
  10.1103/PhysRevLett.76.4500} {\bibfield  {journal} {\bibinfo  {journal}
  {Physical Review Letters}\ }\textbf {\bibinfo {volume} {76}},\ \bibinfo
  {pages} {4500} (\bibinfo {year} {1996})}\BibitemShut {NoStop}%
\bibitem [{\citenamefont {Renn}\ \emph {et~al.}(1997)\citenamefont {Renn},
  \citenamefont {Zozulya}, \citenamefont {Donley}, \citenamefont {Cornell},\
  and\ \citenamefont {Anderson}}]{RZD97}%
  \BibitemOpen
  \bibfield  {author} {\bibinfo {author} {\bibfnamefont {M.~J.}\ \bibnamefont
  {Renn}}, \bibinfo {author} {\bibfnamefont {A.~A.}\ \bibnamefont {Zozulya}},
  \bibinfo {author} {\bibfnamefont {E.~A.}\ \bibnamefont {Donley}}, \bibinfo
  {author} {\bibfnamefont {E.~A.}\ \bibnamefont {Cornell}}, \ and\ \bibinfo
  {author} {\bibfnamefont {D.~Z.}\ \bibnamefont {Anderson}},\ }\href {\doibase
  10.1103/PhysRevA.55.3684} {\bibfield  {journal} {\bibinfo  {journal}
  {Physical Review A}\ }\textbf {\bibinfo {volume} {55}},\ \bibinfo {pages}
  {3684} (\bibinfo {year} {1997})}\BibitemShut {NoStop}%
\bibitem [{\citenamefont {Dall}\ \emph {et~al.}(1999)\citenamefont {Dall},
  \citenamefont {Hoogerland}, \citenamefont {Baldwin},\ and\ \citenamefont
  {Buckman}}]{DHB99}%
  \BibitemOpen
  \bibfield  {author} {\bibinfo {author} {\bibfnamefont {R.~G.}\ \bibnamefont
  {Dall}}, \bibinfo {author} {\bibfnamefont {M.~D.}\ \bibnamefont
  {Hoogerland}}, \bibinfo {author} {\bibfnamefont {K.~G.~H.}\ \bibnamefont
  {Baldwin}}, \ and\ \bibinfo {author} {\bibfnamefont {S.~J.}\ \bibnamefont
  {Buckman}},\ }\href {\doibase 10.1088/1464-4266/1/4/307} {\bibfield
  {journal} {\bibinfo  {journal} {Journal of Optics B: Quantum and
  Semiclassical Optics}\ }\textbf {\bibinfo {volume} {1}},\ \bibinfo {pages}
  {396} (\bibinfo {year} {1999})}\BibitemShut {NoStop}%
\bibitem [{\citenamefont {M{\"{u}}ller}\ \emph {et~al.}(2000)\citenamefont
  {M{\"{u}}ller}, \citenamefont {Cornell}, \citenamefont {Anderson},\ and\
  \citenamefont {Abraham}}]{MCA00}%
  \BibitemOpen
  \bibfield  {author} {\bibinfo {author} {\bibfnamefont {D.}~\bibnamefont
  {M{\"{u}}ller}}, \bibinfo {author} {\bibfnamefont {E.~A.}\ \bibnamefont
  {Cornell}}, \bibinfo {author} {\bibfnamefont {D.~Z.}\ \bibnamefont
  {Anderson}}, \ and\ \bibinfo {author} {\bibfnamefont {E.~R.~I.}\ \bibnamefont
  {Abraham}},\ }\href {\doibase 10.1103/PhysRevA.61.033411} {\bibfield
  {journal} {\bibinfo  {journal} {Physical Review A}\ }\textbf {\bibinfo
  {volume} {61}},\ \bibinfo {pages} {033411} (\bibinfo {year}
  {2000})}\BibitemShut {NoStop}%
\bibitem [{\citenamefont {Christensen}\ \emph {et~al.}(2008)\citenamefont
  {Christensen}, \citenamefont {Will}, \citenamefont {Saba}, \citenamefont
  {Jo}, \citenamefont {Shin}, \citenamefont {Ketterle},\ and\ \citenamefont
  {Pritchard}}]{CWS08}%
  \BibitemOpen
  \bibfield  {author} {\bibinfo {author} {\bibfnamefont {C.~A.}\ \bibnamefont
  {Christensen}}, \bibinfo {author} {\bibfnamefont {S.}~\bibnamefont {Will}},
  \bibinfo {author} {\bibfnamefont {M.}~\bibnamefont {Saba}}, \bibinfo {author}
  {\bibfnamefont {G.-B.}\ \bibnamefont {Jo}}, \bibinfo {author} {\bibfnamefont
  {Y.-I.}\ \bibnamefont {Shin}}, \bibinfo {author} {\bibfnamefont
  {W.}~\bibnamefont {Ketterle}}, \ and\ \bibinfo {author} {\bibfnamefont
  {D.}~\bibnamefont {Pritchard}},\ }\href {\doibase 10.1103/PhysRevA.78.033429}
  {\bibfield  {journal} {\bibinfo  {journal} {Physical Review A}\ }\textbf
  {\bibinfo {volume} {78}},\ \bibinfo {pages} {033429} (\bibinfo {year}
  {2008})}\BibitemShut {NoStop}%
\bibitem [{\citenamefont {Bajcsy}\ \emph {et~al.}(2009)\citenamefont {Bajcsy},
  \citenamefont {Hofferberth}, \citenamefont {Balic}, \citenamefont {Peyronel},
  \citenamefont {Hafezi}, \citenamefont {Zibrov}, \citenamefont
  {Vuleti{\'{c}}},\ and\ \citenamefont {Lukin}}]{BHB09}%
  \BibitemOpen
  \bibfield  {author} {\bibinfo {author} {\bibfnamefont {M.}~\bibnamefont
  {Bajcsy}}, \bibinfo {author} {\bibfnamefont {S.}~\bibnamefont {Hofferberth}},
  \bibinfo {author} {\bibfnamefont {V.}~\bibnamefont {Balic}}, \bibinfo
  {author} {\bibfnamefont {T.}~\bibnamefont {Peyronel}}, \bibinfo {author}
  {\bibfnamefont {M.}~\bibnamefont {Hafezi}}, \bibinfo {author} {\bibfnamefont
  {A.~S.}\ \bibnamefont {Zibrov}}, \bibinfo {author} {\bibfnamefont
  {V.}~\bibnamefont {Vuleti{\'{c}}}}, \ and\ \bibinfo {author} {\bibfnamefont
  {M.~D.}\ \bibnamefont {Lukin}},\ }\href {\doibase
  10.1103/PhysRevLett.102.203902} {\bibfield  {journal} {\bibinfo  {journal}
  {Physical Review Letters}\ }\textbf {\bibinfo {volume} {102}},\ \bibinfo
  {pages} {203902} (\bibinfo {year} {2009})}\BibitemShut {NoStop}%
\bibitem [{\citenamefont {Bajcsy}\ \emph {et~al.}(2011)\citenamefont {Bajcsy},
  \citenamefont {Hofferberth}, \citenamefont {Peyronel}, \citenamefont
  {Bali{\'{c}}}, \citenamefont {Liang}, \citenamefont {Zibrov}, \citenamefont
  {Vuleti{\'{c}}},\ and\ \citenamefont {Lukin}}]{BHP11}%
  \BibitemOpen
  \bibfield  {author} {\bibinfo {author} {\bibfnamefont {M.}~\bibnamefont
  {Bajcsy}}, \bibinfo {author} {\bibfnamefont {S.}~\bibnamefont {Hofferberth}},
  \bibinfo {author} {\bibfnamefont {T.}~\bibnamefont {Peyronel}}, \bibinfo
  {author} {\bibfnamefont {V.}~\bibnamefont {Bali{\'{c}}}}, \bibinfo {author}
  {\bibfnamefont {Q.}~\bibnamefont {Liang}}, \bibinfo {author} {\bibfnamefont
  {A.~S.}\ \bibnamefont {Zibrov}}, \bibinfo {author} {\bibfnamefont
  {V.}~\bibnamefont {Vuleti{\'{c}}}}, \ and\ \bibinfo {author} {\bibfnamefont
  {M.~D.}\ \bibnamefont {Lukin}},\ }\href {\doibase 10.1103/PhysRevA.83.063830}
  {\bibfield  {journal} {\bibinfo  {journal} {Physical Review A}\ }\textbf
  {\bibinfo {volume} {83}},\ \bibinfo {pages} {063830} (\bibinfo {year}
  {2011})}\BibitemShut {NoStop}%
\bibitem [{\citenamefont {Takekoshi}\ and\ \citenamefont {Knize}(2007)}]{TK07}%
  \BibitemOpen
  \bibfield  {author} {\bibinfo {author} {\bibfnamefont {T.}~\bibnamefont
  {Takekoshi}}\ and\ \bibinfo {author} {\bibfnamefont {R.~J.}\ \bibnamefont
  {Knize}},\ }\href {\doibase 10.1103/PhysRevLett.98.210404} {\bibfield
  {journal} {\bibinfo  {journal} {Physical Review Letters}\ }\textbf {\bibinfo
  {volume} {98}},\ \bibinfo {pages} {210404} (\bibinfo {year}
  {2007})}\BibitemShut {NoStop}%
\bibitem [{\citenamefont {Vorrath}\ \emph {et~al.}(2010)\citenamefont
  {Vorrath}, \citenamefont {M{\"{o}}ller}, \citenamefont {Windpassinger},
  \citenamefont {Bongs},\ and\ \citenamefont {Sengstock}}]{VMW10}%
  \BibitemOpen
  \bibfield  {author} {\bibinfo {author} {\bibfnamefont {S.}~\bibnamefont
  {Vorrath}}, \bibinfo {author} {\bibfnamefont {S.~A.}\ \bibnamefont
  {M{\"{o}}ller}}, \bibinfo {author} {\bibfnamefont {P.}~\bibnamefont
  {Windpassinger}}, \bibinfo {author} {\bibfnamefont {K.}~\bibnamefont
  {Bongs}}, \ and\ \bibinfo {author} {\bibfnamefont {K.}~\bibnamefont
  {Sengstock}},\ }\href {\doibase 10.1088/1367-2630/12/12/123015} {\bibfield
  {journal} {\bibinfo  {journal} {New Journal of Physics}\ }\textbf {\bibinfo
  {volume} {12}},\ \bibinfo {pages} {123015} (\bibinfo {year}
  {2010})}\BibitemShut {NoStop}%
\bibitem [{\citenamefont {Peyronel}\ \emph {et~al.}(2012)\citenamefont
  {Peyronel}, \citenamefont {Bajcsy}, \citenamefont {Hofferberth},
  \citenamefont {Balic}, \citenamefont {Hafezi}, \citenamefont {{Qiyu Liang}},
  \citenamefont {Zibrov}, \citenamefont {Vuletic},\ and\ \citenamefont
  {Lukin}}]{PBH12}%
  \BibitemOpen
  \bibfield  {author} {\bibinfo {author} {\bibfnamefont {T.}~\bibnamefont
  {Peyronel}}, \bibinfo {author} {\bibfnamefont {M.}~\bibnamefont {Bajcsy}},
  \bibinfo {author} {\bibfnamefont {S.}~\bibnamefont {Hofferberth}}, \bibinfo
  {author} {\bibfnamefont {V.}~\bibnamefont {Balic}}, \bibinfo {author}
  {\bibfnamefont {M.}~\bibnamefont {Hafezi}}, \bibinfo {author} {\bibnamefont
  {{Qiyu Liang}}}, \bibinfo {author} {\bibfnamefont {A.}~\bibnamefont
  {Zibrov}}, \bibinfo {author} {\bibfnamefont {V.}~\bibnamefont {Vuletic}}, \
  and\ \bibinfo {author} {\bibfnamefont {M.~D.}\ \bibnamefont {Lukin}},\ }\href
  {\doibase 10.1109/JSTQE.2012.2196414} {\bibfield  {journal} {\bibinfo
  {journal} {IEEE Journal of Selected Topics in Quantum Electronics}\ }\textbf
  {\bibinfo {volume} {18}},\ \bibinfo {pages} {1747} (\bibinfo {year}
  {2012})}\BibitemShut {NoStop}%
\bibitem [{\citenamefont {Blatt}\ \emph {et~al.}(2014)\citenamefont {Blatt},
  \citenamefont {Halfmann},\ and\ \citenamefont {Peters}}]{BHP14}%
  \BibitemOpen
  \bibfield  {author} {\bibinfo {author} {\bibfnamefont {F.}~\bibnamefont
  {Blatt}}, \bibinfo {author} {\bibfnamefont {T.}~\bibnamefont {Halfmann}}, \
  and\ \bibinfo {author} {\bibfnamefont {T.}~\bibnamefont {Peters}},\ }\href
  {\doibase 10.1364/OL.39.000446} {\bibfield  {journal} {\bibinfo  {journal}
  {Optics Letters}\ }\textbf {\bibinfo {volume} {39}},\ \bibinfo {pages} {446}
  (\bibinfo {year} {2014})},\ \Eprint {http://arxiv.org/abs/1311.0635v2}
  {arXiv:1311.0635v2} \BibitemShut {NoStop}%
\bibitem [{\citenamefont {Langbecker}\ \emph {et~al.}(2017)\citenamefont
  {Langbecker}, \citenamefont {Noaman}, \citenamefont {Kj{\ae}rgaard},
  \citenamefont {Benabid},\ and\ \citenamefont {Windpassinger}}]{LNK17}%
  \BibitemOpen
  \bibfield  {author} {\bibinfo {author} {\bibfnamefont {M.}~\bibnamefont
  {Langbecker}}, \bibinfo {author} {\bibfnamefont {M.}~\bibnamefont {Noaman}},
  \bibinfo {author} {\bibfnamefont {N.}~\bibnamefont {Kj{\ae}rgaard}}, \bibinfo
  {author} {\bibfnamefont {F.}~\bibnamefont {Benabid}}, \ and\ \bibinfo
  {author} {\bibfnamefont {P.}~\bibnamefont {Windpassinger}},\ }\href {\doibase
  10.1103/PhysRevA.96.041402} {\bibfield  {journal} {\bibinfo  {journal}
  {Physical Review A}\ }\textbf {\bibinfo {volume} {96}},\ \bibinfo {pages}
  {041402(R)} (\bibinfo {year} {2017})}\BibitemShut {NoStop}%
\bibitem [{\citenamefont {Okaba}\ \emph {et~al.}(2014)\citenamefont {Okaba},
  \citenamefont {Takano}, \citenamefont {Benabid}, \citenamefont {Bradley},
  \citenamefont {Vincetti}, \citenamefont {Maizelis}, \citenamefont
  {Yampol'skii}, \citenamefont {Nori},\ and\ \citenamefont {Katori}}]{OTB14}%
  \BibitemOpen
  \bibfield  {author} {\bibinfo {author} {\bibfnamefont {S.}~\bibnamefont
  {Okaba}}, \bibinfo {author} {\bibfnamefont {T.}~\bibnamefont {Takano}},
  \bibinfo {author} {\bibfnamefont {F.}~\bibnamefont {Benabid}}, \bibinfo
  {author} {\bibfnamefont {T.}~\bibnamefont {Bradley}}, \bibinfo {author}
  {\bibfnamefont {L.}~\bibnamefont {Vincetti}}, \bibinfo {author}
  {\bibfnamefont {Z.}~\bibnamefont {Maizelis}}, \bibinfo {author}
  {\bibfnamefont {V.}~\bibnamefont {Yampol'skii}}, \bibinfo {author}
  {\bibfnamefont {F.}~\bibnamefont {Nori}}, \ and\ \bibinfo {author}
  {\bibfnamefont {H.}~\bibnamefont {Katori}},\ }\href {\doibase
  10.1038/ncomms5096} {\bibfield  {journal} {\bibinfo  {journal} {Nature
  Communications}\ }\textbf {\bibinfo {volume} {5}},\ \bibinfo {pages} {4096}
  (\bibinfo {year} {2014})},\ \Eprint {http://arxiv.org/abs/1408.0659}
  {arXiv:1408.0659} \BibitemShut {NoStop}%
\bibitem [{\citenamefont {Langbecker}\ \emph {et~al.}(2018)\citenamefont
  {Langbecker}, \citenamefont {Wirtz}, \citenamefont {Knoch}, \citenamefont
  {Noaman}, \citenamefont {Speck},\ and\ \citenamefont
  {Windpassinger}}]{LWK18}%
  \BibitemOpen
  \bibfield  {author} {\bibinfo {author} {\bibfnamefont {M.}~\bibnamefont
  {Langbecker}}, \bibinfo {author} {\bibfnamefont {R.}~\bibnamefont {Wirtz}},
  \bibinfo {author} {\bibfnamefont {F.}~\bibnamefont {Knoch}}, \bibinfo
  {author} {\bibfnamefont {M.}~\bibnamefont {Noaman}}, \bibinfo {author}
  {\bibfnamefont {T.}~\bibnamefont {Speck}}, \ and\ \bibinfo {author}
  {\bibfnamefont {P.}~\bibnamefont {Windpassinger}},\ }\href {\doibase
  10.1088/1367-2630/aad9bb} {\bibfield  {journal} {\bibinfo  {journal} {New
  Journal of Physics}\ }\textbf {\bibinfo {volume} {20}},\ \bibinfo {pages}
  {083038} (\bibinfo {year} {2018})}\BibitemShut {NoStop}%
\bibitem [{\citenamefont {Hilton}\ \emph {et~al.}(2019)\citenamefont {Hilton},
  \citenamefont {Perrella}, \citenamefont {Luiten},\ and\ \citenamefont
  {Light}}]{HPL19}%
  \BibitemOpen
  \bibfield  {author} {\bibinfo {author} {\bibfnamefont {A.~P.}\ \bibnamefont
  {Hilton}}, \bibinfo {author} {\bibfnamefont {C.}~\bibnamefont {Perrella}},
  \bibinfo {author} {\bibfnamefont {A.~N.}~\bibnamefont {Luiten}}, \ and\ \bibinfo
  {author} {\bibfnamefont {P.~S.}~\bibnamefont {Light}},\ }\href {\doibase
  10.1103/PhysRevApplied.11.024065} {\bibfield  {journal} {\bibinfo  {journal}
  {Physical Review Applied}\ }\textbf {\bibinfo {volume} {11}},\ \bibinfo
  {pages} {024065} (\bibinfo {year} {2019})}\BibitemShut {NoStop}%
\bibitem [{\citenamefont {Hilton}\ \emph {et~al.}(2018)\citenamefont {Hilton},
  \citenamefont {Perrella}, \citenamefont {Benabid}, \citenamefont {Sparkes},
  \citenamefont {Luiten},\ and\ \citenamefont {Light}}]{HPB18}%
  \BibitemOpen
  \bibfield  {author} {\bibinfo {author} {\bibfnamefont {A.~P.}\ \bibnamefont
  {Hilton}}, \bibinfo {author} {\bibfnamefont {C.}~\bibnamefont {Perrella}},
  \bibinfo {author} {\bibfnamefont {F.}~\bibnamefont {Benabid}}, \bibinfo
  {author} {\bibfnamefont {B.~M.}\ \bibnamefont {Sparkes}}, \bibinfo {author}
  {\bibfnamefont {A.~N.}\ \bibnamefont {Luiten}}, \ and\ \bibinfo {author}
  {\bibfnamefont {P.~S.}~\bibnamefont {Light}},\ }\href {\doibase
  10.1103/PhysRevApplied.10.044034} {\bibfield  {journal} {\bibinfo  {journal}
  {Physical Review Applied}\ }\textbf {\bibinfo {volume} {10}},\ \bibinfo
  {pages} {044034} (\bibinfo {year} {2018})}\BibitemShut {NoStop}%
\bibitem [{\citenamefont {Xin}\ \emph {et~al.}(2018)\citenamefont {Xin},
  \citenamefont {Leong}, \citenamefont {Chen},\ and\ \citenamefont
  {Lan}}]{XLC18}%
  \BibitemOpen
  \bibfield  {author} {\bibinfo {author} {\bibfnamefont {M.}~\bibnamefont
  {Xin}}, \bibinfo {author} {\bibfnamefont {W.~S.}\ \bibnamefont {Leong}},
  \bibinfo {author} {\bibfnamefont {Z.}~\bibnamefont {Chen}}, \ and\ \bibinfo
  {author} {\bibfnamefont {S.-Y.}\ \bibnamefont {Lan}},\ }\href {\doibase
  10.1126/sciadv.1701723} {\bibfield  {journal} {\bibinfo  {journal} {Science
  Advances}\ }\textbf {\bibinfo {volume} {4}},\ \bibinfo {pages} {e1701723}
  (\bibinfo {year} {2018})},\ \Eprint {http://arxiv.org/abs/1705.08062}
  {arXiv:1705.08062} \BibitemShut {NoStop}%
\bibitem [{\citenamefont {Noaman}\ \emph {et~al.}(2018)\citenamefont {Noaman},
  \citenamefont {Langbecker},\ and\ \citenamefont {Windpassinger}}]{NLW18}%
  \BibitemOpen
  \bibfield  {author} {\bibinfo {author} {\bibfnamefont {M.}~\bibnamefont
  {Noaman}}, \bibinfo {author} {\bibfnamefont {M.}~\bibnamefont {Langbecker}},
  \ and\ \bibinfo {author} {\bibfnamefont {P.}~\bibnamefont {Windpassinger}},\
  }\href {\doibase 10.1364/OL.43.003925} {\bibfield  {journal} {\bibinfo
  {journal} {Optics Letters}\ }\textbf {\bibinfo {volume} {43}},\ \bibinfo
  {pages} {3925} (\bibinfo {year} {2018})}\BibitemShut {NoStop}%
\bibitem [{\citenamefont {Xin}\ \emph {et~al.}(2019)\citenamefont {Xin},
  \citenamefont {Leong}, \citenamefont {Chen},\ and\ \citenamefont
  {Lan}}]{XLC19}%
  \BibitemOpen
  \bibfield  {author} {\bibinfo {author} {\bibfnamefont {M.}~\bibnamefont
  {Xin}}, \bibinfo {author} {\bibfnamefont {W.~S.}\ \bibnamefont {Leong}},
  \bibinfo {author} {\bibfnamefont {Z.}~\bibnamefont {Chen}}, \ and\ \bibinfo
  {author} {\bibfnamefont {S.-Y.}\ \bibnamefont {Lan}},\ }\href {\doibase
  10.1103/PhysRevLett.122.163901} {\bibfield  {journal} {\bibinfo  {journal}
  {Physical Review Letters}\ }\textbf {\bibinfo {volume} {122}},\ \bibinfo
  {pages} {163901} (\bibinfo {year} {2019})}\BibitemShut {NoStop}%
\bibitem [{\citenamefont {Yoon}\ and\ \citenamefont {Bajcsy}(2019)}]{YB19}%
  \BibitemOpen
  \bibfield  {author} {\bibinfo {author} {\bibfnamefont {T.}~\bibnamefont
  {Yoon}}\ and\ \bibinfo {author} {\bibfnamefont {M.}~\bibnamefont {Bajcsy}},\
  }\href {\doibase 10.1103/PhysRevA.99.023415} {\bibfield  {journal} {\bibinfo
  {journal} {Physical Review A}\ }\textbf {\bibinfo {volume} {99}},\ \bibinfo
  {pages} {023415} (\bibinfo {year} {2019})}\BibitemShut {NoStop}%
\bibitem [{\citenamefont {Yoon}\ \emph {et~al.}(2019)\citenamefont {Yoon},
  \citenamefont {Ding}, \citenamefont {Flannery}, \citenamefont {Rajabi},\ and\
  \citenamefont {Bajcsy}}]{YDF19}%
  \BibitemOpen
  \bibfield  {author} {\bibinfo {author} {\bibfnamefont {T.}~\bibnamefont
  {Yoon}}, \bibinfo {author} {\bibfnamefont {Z.}~\bibnamefont {Ding}}, \bibinfo
  {author} {\bibfnamefont {J.}~\bibnamefont {Flannery}}, \bibinfo {author}
  {\bibfnamefont {F.}~\bibnamefont {Rajabi}}, \ and\ \bibinfo {author}
  {\bibfnamefont {M.}~\bibnamefont {Bajcsy}},\ }\href {\doibase
  10.1364/OE.27.017592} {\bibfield  {journal} {\bibinfo  {journal} {Optics
  Express}\ }\textbf {\bibinfo {volume} {27}},\ \bibinfo {pages} {17592}
  (\bibinfo {year} {2019})}\BibitemShut {NoStop}%
\bibitem [{\citenamefont {Okaba}\ \emph {et~al.}(2019)\citenamefont {Okaba},
  \citenamefont {Yu}, \citenamefont {Vincetti}, \citenamefont {Benabid},\ and\
  \citenamefont {Katori}}]{OYV19}%
  \BibitemOpen
  \bibfield  {author} {\bibinfo {author} {\bibfnamefont {S.}~\bibnamefont
  {Okaba}}, \bibinfo {author} {\bibfnamefont {D.}~\bibnamefont {Yu}}, \bibinfo
  {author} {\bibfnamefont {L.}~\bibnamefont {Vincetti}}, \bibinfo {author}
  {\bibfnamefont {F.}~\bibnamefont {Benabid}}, \ and\ \bibinfo {author}
  {\bibfnamefont {H.}~\bibnamefont {Katori}},\ }\href {\doibase
  10.1038/s42005-019-0237-2} {\bibfield  {journal} {\bibinfo  {journal}
  {Communications Physics}\ }\textbf {\bibinfo {volume} {2}},\ \bibinfo {pages}
  {136} (\bibinfo {year} {2019})}\BibitemShut {NoStop}%
\bibitem [{\citenamefont {Hilton}\ \emph {et~al.}(2020)\citenamefont {Hilton},
  \citenamefont {Luiten},\ and\ \citenamefont {Light}}]{HLL20}%
  \BibitemOpen
  \bibfield  {author} {\bibinfo {author} {\bibfnamefont {A.~P.}\ \bibnamefont
  {Hilton}}, \bibinfo {author} {\bibfnamefont {A.~N.}\ \bibnamefont {Luiten}},
  \ and\ \bibinfo {author} {\bibfnamefont {P.~S.}\ \bibnamefont {Light}},\
  }\href {\doibase 10.1088/1367-2630/ab753a} {\bibfield  {journal} {\bibinfo
  {journal} {New Journal of Physics}\ }\textbf {\bibinfo {volume} {22}},\
  \bibinfo {pages} {033042} (\bibinfo {year} {2020})},\ \Eprint
  {http://arxiv.org/abs/1911.02708} {arXiv:1911.02708} \BibitemShut {NoStop}%
\bibitem [{\citenamefont {Wang}\ \emph {et~al.}(2020)\citenamefont {Wang},
  \citenamefont {Chai}, \citenamefont {Xin}, \citenamefont {Leong},
  \citenamefont {Chen},\ and\ \citenamefont {Lan}}]{WCX20}%
  \BibitemOpen
  \bibfield  {author} {\bibinfo {author} {\bibfnamefont {Y.}~\bibnamefont
  {Wang}}, \bibinfo {author} {\bibfnamefont {S.}~\bibnamefont {Chai}}, \bibinfo
  {author} {\bibfnamefont {M.}~\bibnamefont {Xin}}, \bibinfo {author}
  {\bibfnamefont {W.~S.}\ \bibnamefont {Leong}}, \bibinfo {author}
  {\bibfnamefont {Z.}~\bibnamefont {Chen}}, \ and\ \bibinfo {author}
  {\bibfnamefont {S.-Y.}\ \bibnamefont {Lan}},\ }\href {\doibase
  10.3390/FIB8050028} {\bibfield  {journal} {\bibinfo  {journal} {Fibers}\
  }\textbf {\bibinfo {volume} {8}} (\bibinfo {year} {2020}),\
  10.3390/FIB8050028}\BibitemShut {NoStop}%
\bibitem [{\citenamefont {Leong}\ \emph
  {et~al.}(2020{\natexlab{a}})\citenamefont {Leong}, \citenamefont {Xin},
  \citenamefont {Huang}, \citenamefont {Chen},\ and\ \citenamefont
  {Lan}}]{LXH20}%
  \BibitemOpen
  \bibfield  {author} {\bibinfo {author} {\bibfnamefont {W.~S.}\ \bibnamefont
  {Leong}}, \bibinfo {author} {\bibfnamefont {M.}~\bibnamefont {Xin}}, \bibinfo
  {author} {\bibfnamefont {C.}~\bibnamefont {Huang}}, \bibinfo {author}
  {\bibfnamefont {Z.}~\bibnamefont {Chen}}, \ and\ \bibinfo {author}
  {\bibfnamefont {S.-y.}\ \bibnamefont {Lan}},\ }\href {\doibase
  10.1103/PhysRevResearch.2.033320} {\bibfield  {journal} {\bibinfo  {journal}
  {Physical Review Research}\ }\textbf {\bibinfo {volume} {2}},\ \bibinfo
  {pages} {033320} (\bibinfo {year} {2020}{\natexlab{a}})}\BibitemShut
  {NoStop}%
\bibitem [{\citenamefont {Li}\ \emph {et~al.}(2020)\citenamefont {Li},
  \citenamefont {Islam},\ and\ \citenamefont {Windpassinger}}]{LIW20}%
  \BibitemOpen
  \bibfield  {author} {\bibinfo {author} {\bibfnamefont {W.}~\bibnamefont
  {Li}}, \bibinfo {author} {\bibfnamefont {P.}~\bibnamefont {Islam}}, \ and\
  \bibinfo {author} {\bibfnamefont {P.}~\bibnamefont {Windpassinger}},\ }\href
  {\doibase 10.1103/PhysRevLett.125.150501} {\bibfield  {journal} {\bibinfo
  {journal} {Physical Review Letters}\ }\textbf {\bibinfo {volume} {125}},\
  \bibinfo {pages} {150501} (\bibinfo {year} {2020})}\BibitemShut {NoStop}%
\bibitem [{\citenamefont {Leong}\ \emph
  {et~al.}(2020{\natexlab{b}})\citenamefont {Leong}, \citenamefont {Xin},
  \citenamefont {Chen}, \citenamefont {Chai}, \citenamefont {Wang},\ and\
  \citenamefont {Lan}}]{LXC20}%
  \BibitemOpen
  \bibfield  {author} {\bibinfo {author} {\bibfnamefont {W.~S.}\ \bibnamefont
  {Leong}}, \bibinfo {author} {\bibfnamefont {M.}~\bibnamefont {Xin}}, \bibinfo
  {author} {\bibfnamefont {Z.}~\bibnamefont {Chen}}, \bibinfo {author}
  {\bibfnamefont {S.}~\bibnamefont {Chai}}, \bibinfo {author} {\bibfnamefont
  {Y.}~\bibnamefont {Wang}}, \ and\ \bibinfo {author} {\bibfnamefont {S.-Y.}\
  \bibnamefont {Lan}},\ }\href {\doibase 10.1038/s41467-020-19030-2} {\bibfield
   {journal} {\bibinfo  {journal} {Nature Communications}\ }\textbf {\bibinfo
  {volume} {11}},\ \bibinfo {pages} {5295} (\bibinfo {year}
  {2020}{\natexlab{b}})}\BibitemShut {NoStop}%
\bibitem [{\citenamefont {Noh}\ and\ \citenamefont {Angelakis}(2017)}]{NA17}%
  \BibitemOpen
  \bibfield  {author} {\bibinfo {author} {\bibfnamefont {C.}~\bibnamefont
  {Noh}}\ and\ \bibinfo {author} {\bibfnamefont {D.~G.}\ \bibnamefont
  {Angelakis}},\ }\href {\doibase 10.1088/0034-4885/80/1/016401} {\bibfield
  {journal} {\bibinfo  {journal} {Reports on Progress in Physics}\ }\textbf
  {\bibinfo {volume} {80}},\ \bibinfo {pages} {016401} (\bibinfo {year}
  {2017})},\ \Eprint {http://arxiv.org/abs/1604.04433} {arXiv:1604.04433}
  \BibitemShut {NoStop}%
\bibitem [{\citenamefont {Grimm}\ \emph {et~al.}(2000)\citenamefont {Grimm},
  \citenamefont {Weidem{\"{u}}ller},\ and\ \citenamefont
  {Ovchinnikov}}]{GWO00}%
  \BibitemOpen
  \bibfield  {author} {\bibinfo {author} {\bibfnamefont {R.}~\bibnamefont
  {Grimm}}, \bibinfo {author} {\bibfnamefont {M.}~\bibnamefont
  {Weidem{\"{u}}ller}}, \ and\ \bibinfo {author} {\bibfnamefont {Y.~B.}\
  \bibnamefont {Ovchinnikov}},\ }\href {\doibase 10.1016/S1049-250X(08)60186-X}
  {\bibfield  {journal} {\bibinfo  {journal} {Advances In Atomic, Molecular,
  and Optical Physics}\ }\textbf {\bibinfo {volume} {42}},\ \bibinfo {pages}
  {95} (\bibinfo {year} {2000})}\BibitemShut {NoStop}%
\bibitem [{\citenamefont {Russell}(2003)}]{R03}%
  \BibitemOpen
  \bibfield  {author} {\bibinfo {author} {\bibfnamefont {P.}~\bibnamefont
  {Russell}},\ }\href {\doibase 10.1126/science.1079280} {\bibfield  {journal}
  {\bibinfo  {journal} {Science (New York, N.Y.)}\ }\textbf {\bibinfo {volume}
  {299}},\ \bibinfo {pages} {358} (\bibinfo {year} {2003})}\BibitemShut
  {NoStop}%
\bibitem [{\citenamefont {Couny}\ \emph {et~al.}(2006)\citenamefont {Couny},
  \citenamefont {Benabid},\ and\ \citenamefont {Light}}]{CBL06}%
  \BibitemOpen
  \bibfield  {author} {\bibinfo {author} {\bibfnamefont {F.}~\bibnamefont
  {Couny}}, \bibinfo {author} {\bibfnamefont {F.}~\bibnamefont {Benabid}}, \
  and\ \bibinfo {author} {\bibfnamefont {P.~S.}\ \bibnamefont {Light}},\ }\href
  {\doibase 10.1364/OL.31.003574} {\bibfield  {journal} {\bibinfo  {journal}
  {Optics Letters}\ }\textbf {\bibinfo {volume} {31}},\ \bibinfo {pages} {3574}
  (\bibinfo {year} {2006})}\BibitemShut {NoStop}%
\bibitem [{\citenamefont {Cregan}\ \emph {et~al.}(1999)\citenamefont {Cregan},
  \citenamefont {Mangan}, \citenamefont {Knight}, \citenamefont {Birks},
  \citenamefont {Russell}, \citenamefont {Roberts},\ and\ \citenamefont
  {Allan}}]{CMK99}%
  \BibitemOpen
  \bibfield  {author} {\bibinfo {author} {\bibfnamefont {R.~F.}\ \bibnamefont
  {Cregan}}, \bibinfo {author} {\bibfnamefont {B.~J.}\ \bibnamefont {Mangan}},
  \bibinfo {author} {\bibfnamefont {J.~C.}\ \bibnamefont {Knight}}, \bibinfo
  {author} {\bibfnamefont {T.~A.}\ \bibnamefont {Birks}}, \bibinfo {author}
  {\bibfnamefont {P.~S.~J.}\ \bibnamefont {Russell}}, \bibinfo {author}
  {\bibfnamefont {P.~J.}\ \bibnamefont {Roberts}}, \ and\ \bibinfo {author}
  {\bibfnamefont {D.~C.}\ \bibnamefont {Allan}},\ }\href {\doibase
  10.1126/science.285.5433.1537} {\bibfield  {journal} {\bibinfo  {journal}
  {Science}\ }\textbf {\bibinfo {volume} {285}},\ \bibinfo {pages} {1537}
  (\bibinfo {year} {1999})}\BibitemShut {NoStop}%
\bibitem [{\citenamefont {Sulzbach}\ \emph {et~al.}(2019)\citenamefont
  {Sulzbach}, \citenamefont {Peters},\ and\ \citenamefont {Walser}}]{SPW19}%
  \BibitemOpen
  \bibfield  {author} {\bibinfo {author} {\bibfnamefont {R.}~\bibnamefont
  {Sulzbach}}, \bibinfo {author} {\bibfnamefont {T.}~\bibnamefont {Peters}}, \
  and\ \bibinfo {author} {\bibfnamefont {R.}~\bibnamefont {Walser}},\ }\href
  {\doibase 10.1103/PhysRevA.100.013847} {\bibfield  {journal} {\bibinfo
  {journal} {Physical Review A}\ }\textbf {\bibinfo {volume} {100}},\ \bibinfo
  {pages} {013847} (\bibinfo {year} {2019})}\BibitemShut {NoStop}%
\bibitem [{\citenamefont {Wegmuller}\ \emph {et~al.}(2005)\citenamefont
  {Wegmuller}, \citenamefont {Legr{\'{e}}}, \citenamefont {Gisin},
  \citenamefont {Hansen}, \citenamefont {Jakobsen},\ and\ \citenamefont
  {Broeng}}]{WLG05}%
  \BibitemOpen
  \bibfield  {author} {\bibinfo {author} {\bibfnamefont {M.}~\bibnamefont
  {Wegmuller}}, \bibinfo {author} {\bibfnamefont {M.}~\bibnamefont
  {Legr{\'{e}}}}, \bibinfo {author} {\bibfnamefont {N.}~\bibnamefont {Gisin}},
  \bibinfo {author} {\bibfnamefont {T.}~\bibnamefont {Hansen}}, \bibinfo
  {author} {\bibfnamefont {C.}~\bibnamefont {Jakobsen}}, \ and\ \bibinfo
  {author} {\bibfnamefont {J.}~\bibnamefont {Broeng}},\ }\href
  {http://www.ncbi.nlm.nih.gov/pubmed/19495021} {\bibfield  {journal} {\bibinfo
   {journal} {Optics Express}\ }\textbf {\bibinfo {volume} {13}},\ \bibinfo
  {pages} {1457} (\bibinfo {year} {2005})}\BibitemShut {NoStop}%
\bibitem [{\citenamefont {Blatt}\ \emph {et~al.}(2016)\citenamefont {Blatt},
  \citenamefont {Simeonov}, \citenamefont {Halfmann},\ and\ \citenamefont
  {Peters}}]{BSH16}%
  \BibitemOpen
  \bibfield  {author} {\bibinfo {author} {\bibfnamefont {F.}~\bibnamefont
  {Blatt}}, \bibinfo {author} {\bibfnamefont {L.~S.}\ \bibnamefont {Simeonov}},
  \bibinfo {author} {\bibfnamefont {T.}~\bibnamefont {Halfmann}}, \ and\
  \bibinfo {author} {\bibfnamefont {T.}~\bibnamefont {Peters}},\ }\href
  {\doibase 10.1103/PhysRevA.94.043833} {\bibfield  {journal} {\bibinfo
  {journal} {Physical Review A}\ }\textbf {\bibinfo {volume} {94}},\ \bibinfo
  {pages} {043833} (\bibinfo {year} {2016})}\BibitemShut {NoStop}%
\bibitem [{\citenamefont {Peters}\ \emph {et~al.}(2020)\citenamefont {Peters},
  \citenamefont {Wang}, \citenamefont {Neumann}, \citenamefont {Simeonov},\
  and\ \citenamefont {Halfmann}}]{PWN20}%
  \BibitemOpen
  \bibfield  {author} {\bibinfo {author} {\bibfnamefont {T.}~\bibnamefont
  {Peters}}, \bibinfo {author} {\bibfnamefont {T.-P.}\ \bibnamefont {Wang}},
  \bibinfo {author} {\bibfnamefont {A.}~\bibnamefont {Neumann}}, \bibinfo
  {author} {\bibfnamefont {L.~S.}\ \bibnamefont {Simeonov}}, \ and\ \bibinfo
  {author} {\bibfnamefont {T.}~\bibnamefont {Halfmann}},\ }\href {\doibase
  10.1364/OE.383999} {\bibfield  {journal} {\bibinfo  {journal} {Optics
  Express}\ }\textbf {\bibinfo {volume} {28}},\ \bibinfo {pages} {5340}
  (\bibinfo {year} {2020})}\BibitemShut {NoStop}%
\bibitem [{\citenamefont {Chu}\ \emph {et~al.}(1986)\citenamefont {Chu},
  \citenamefont {Bjorkholm}, \citenamefont {Ashkin},\ and\ \citenamefont
  {Cable}}]{CBA86}%
  \BibitemOpen
  \bibfield  {author} {\bibinfo {author} {\bibfnamefont {S.}~\bibnamefont
  {Chu}}, \bibinfo {author} {\bibfnamefont {J.~E.}\ \bibnamefont {Bjorkholm}},
  \bibinfo {author} {\bibfnamefont {A.}~\bibnamefont {Ashkin}}, \ and\ \bibinfo
  {author} {\bibfnamefont {A.}~\bibnamefont {Cable}},\ }\href {\doibase
  10.1103/PhysRevLett.57.314} {\bibfield  {journal} {\bibinfo  {journal} {Phys.
  Rev. Lett.}\ }\textbf {\bibinfo {volume} {57}},\ \bibinfo {pages} {314}
  (\bibinfo {year} {1986})}\BibitemShut {NoStop}%
\bibitem [{\citenamefont {Miller}\ \emph {et~al.}(1993)\citenamefont {Miller},
  \citenamefont {Cline},\ and\ \citenamefont {Heinzen}}]{MCH93a}%
  \BibitemOpen
  \bibfield  {author} {\bibinfo {author} {\bibfnamefont {J.~D.}\ \bibnamefont
  {Miller}}, \bibinfo {author} {\bibfnamefont {R.~A.}\ \bibnamefont {Cline}}, \
  and\ \bibinfo {author} {\bibfnamefont {D.~J.}\ \bibnamefont {Heinzen}},\
  }\href {\doibase 10.1103/PhysRevA.47.R4567} {\bibfield  {journal} {\bibinfo
  {journal} {Physical Review A}\ }\textbf {\bibinfo {volume} {47}},\ \bibinfo
  {pages} {R4567} (\bibinfo {year} {1993})}\BibitemShut {NoStop}%
\bibitem [{\citenamefont {Kuppens}\ \emph {et~al.}(2000)\citenamefont
  {Kuppens}, \citenamefont {Corwin}, \citenamefont {Miller}, \citenamefont
  {Chupp},\ and\ \citenamefont {Wieman}}]{KCM00}%
  \BibitemOpen
  \bibfield  {author} {\bibinfo {author} {\bibfnamefont {S.~J.~M.}\
  \bibnamefont {Kuppens}}, \bibinfo {author} {\bibfnamefont {K.~L.}\
  \bibnamefont {Corwin}}, \bibinfo {author} {\bibfnamefont {K.~W.}\
  \bibnamefont {Miller}}, \bibinfo {author} {\bibfnamefont {T.~E.}\
  \bibnamefont {Chupp}}, \ and\ \bibinfo {author} {\bibfnamefont {C.~E.}\
  \bibnamefont {Wieman}},\ }\href {\doibase 10.1103/PhysRevA.62.013406}
  {\bibfield  {journal} {\bibinfo  {journal} {Physical Review A}\ }\textbf
  {\bibinfo {volume} {62}},\ \bibinfo {pages} {013406} (\bibinfo {year}
  {2000})}\BibitemShut {NoStop}%
\bibitem [{\citenamefont {Anderson}\ \emph {et~al.}(1994)\citenamefont
  {Anderson}, \citenamefont {Petrich}, \citenamefont {Ensher},\ and\
  \citenamefont {Cornell}}]{APE94}%
  \BibitemOpen
  \bibfield  {author} {\bibinfo {author} {\bibfnamefont {M.~H.}~\bibnamefont
  {Anderson}}, \bibinfo {author} {\bibfnamefont {W.}~\bibnamefont {Petrich}},
  \bibinfo {author} {\bibfnamefont {J.~R.}~\bibnamefont {Ensher}}, \ and\ \bibinfo
  {author} {\bibfnamefont {E.~A.}~\bibnamefont {Cornell}},\ }\href {\doibase
  10.1103/PhysRevA.50.R3597} {\bibfield  {journal} {\bibinfo  {journal}
  {Physical Review A}\ }\textbf {\bibinfo {volume} {50}},\ \bibinfo {pages}
  {R3597} (\bibinfo {year} {1994})}\BibitemShut {NoStop}%
\bibitem [{\citenamefont {Harris}(1997)}]{H97}%
  \BibitemOpen
  \bibfield  {author} {\bibinfo {author} {\bibfnamefont {S.~E.}\ \bibnamefont
  {Harris}},\ }\href@noop {} {\bibfield  {journal} {\bibinfo  {journal}
  {Physics Today}\ }\textbf {\bibinfo {volume} {50}},\ \bibinfo {pages} {36}
  (\bibinfo {year} {1997})}\BibitemShut {NoStop}%
\bibitem [{\citenamefont {Fleischhauer}\ \emph {et~al.}(2005)\citenamefont
  {Fleischhauer}, \citenamefont {Imamoǧlu},\ and\ \citenamefont
  {Marangos}}]{FIM05}%
  \BibitemOpen
  \bibfield  {author} {\bibinfo {author} {\bibfnamefont {M.}~\bibnamefont
  {Fleischhauer}}, \bibinfo {author} {\bibfnamefont {A.}~\bibnamefont
  {Imamoglu}}, \ and\ \bibinfo {author} {\bibfnamefont {J.~P.}\ \bibnamefont
  {Marangos}},\ }\href {\doibase 10.1103/RevModPhys.77.633} {\bibfield
  {journal} {\bibinfo  {journal} {Reviews of Modern Physics}\ }\textbf
  {\bibinfo {volume} {77}},\ \bibinfo {pages} {633} (\bibinfo {year}
  {2005})}\BibitemShut {NoStop}%
\bibitem [{\citenamefont {Perrella}\ \emph {et~al.}(2012)\citenamefont
  {Perrella}, \citenamefont {Light}, \citenamefont {Stace}, \citenamefont
  {Benabid},\ and\ \citenamefont {Luiten}}]{PLS12}%
  \BibitemOpen
  \bibfield  {author} {\bibinfo {author} {\bibfnamefont {C.}~\bibnamefont
  {Perrella}}, \bibinfo {author} {\bibfnamefont {P.~S.}~\bibnamefont {Light}},
  \bibinfo {author} {\bibfnamefont {T.~M.}~\bibnamefont {Stace}}, \bibinfo
  {author} {\bibfnamefont {F.}~\bibnamefont {Benabid}}, \ and\ \bibinfo
  {author} {\bibfnamefont {A.~N.}~\bibnamefont {Luiten}},\ }\href {\doibase
  10.1103/PhysRevA.85.012518} {\bibfield  {journal} {\bibinfo  {journal}
  {Physical Review A}\ }\textbf {\bibinfo {volume} {85}},\ \bibinfo {pages}
  {012518} (\bibinfo {year} {2012})}\BibitemShut {NoStop}%
\bibitem [{\citenamefont {Peters}\ \emph {et~al.}(2012)\citenamefont {Peters},
  \citenamefont {Wittrock}, \citenamefont {Blatt}, \citenamefont {Halfmann},\
  and\ \citenamefont {Yatsenko}}]{PWB12}%
  \BibitemOpen
  \bibfield  {author} {\bibinfo {author} {\bibfnamefont {T.}~\bibnamefont
  {Peters}}, \bibinfo {author} {\bibfnamefont {B.}~\bibnamefont {Wittrock}},
  \bibinfo {author} {\bibfnamefont {F.}~\bibnamefont {Blatt}}, \bibinfo
  {author} {\bibfnamefont {T.}~\bibnamefont {Halfmann}}, \ and\ \bibinfo
  {author} {\bibfnamefont {L.~P.}~\bibnamefont {Yatsenko}},\ }\href {\doibase
  10.1103/PhysRevA.85.063416} {\bibfield  {journal} {\bibinfo  {journal}
  {Physical Review A}\ }\textbf {\bibinfo {volume} {85}},\ \bibinfo {pages}
  {063416} (\bibinfo {year} {2012})}\BibitemShut {NoStop}%
\bibitem [{\citenamefont {Ghosh}\ \emph {et~al.}(2006)\citenamefont {Ghosh},
  \citenamefont {Bhagwat}, \citenamefont {Renshaw}, \citenamefont {Goh},
  \citenamefont {Gaeta},\ and\ \citenamefont {Kirby}}]{GBR06}%
  \BibitemOpen
  \bibfield  {author} {\bibinfo {author} {\bibfnamefont {S.}~\bibnamefont
  {Ghosh}}, \bibinfo {author} {\bibfnamefont {A.~R.}\ \bibnamefont {Bhagwat}},
  \bibinfo {author} {\bibfnamefont {C.~K.}\ \bibnamefont {Renshaw}}, \bibinfo
  {author} {\bibfnamefont {S.}~\bibnamefont {Goh}}, \bibinfo {author}
  {\bibfnamefont {A.~L.}\ \bibnamefont {Gaeta}}, \ and\ \bibinfo {author}
  {\bibfnamefont {B.~J.}\ \bibnamefont {Kirby}},\ }\href {\doibase
  10.1103/PhysRevLett.97.023603} {\bibfield  {journal} {\bibinfo  {journal}
  {Physical Review Letters}\ }\textbf {\bibinfo {volume} {97}},\ \bibinfo
  {pages} {023603} (\bibinfo {year} {2006})}\BibitemShut {NoStop}%
\bibitem [{\citenamefont {Barrett}\ \emph {et~al.}(2001)\citenamefont
  {Barrett}, \citenamefont {Sauer},\ and\ \citenamefont {Chapman}}]{BSC01}%
  \BibitemOpen
  \bibfield  {author} {\bibinfo {author} {\bibfnamefont {M.~D.}\ \bibnamefont
  {Barrett}}, \bibinfo {author} {\bibfnamefont {J.~A.}\ \bibnamefont {Sauer}},
  \ and\ \bibinfo {author} {\bibfnamefont {M.~S.}\ \bibnamefont {Chapman}},\
  }\href {\doibase 10.1103/PhysRevLett.87.010404} {\bibfield  {journal}
  {\bibinfo  {journal} {Physical Review Letters}\ }\textbf {\bibinfo {volume}
  {87}},\ \bibinfo {pages} {010404} (\bibinfo {year} {2001})}\BibitemShut
  {NoStop}%
\bibitem [{\citenamefont {Ketterle}\ and\ \citenamefont {van
  Druten}(1996)}]{KV96}%
  \BibitemOpen
  \bibfield  {author} {\bibinfo {author} {\bibfnamefont {W.}~\bibnamefont
  {Ketterle}}\ and\ \bibinfo {author} {\bibfnamefont {N.}~\bibnamefont {van
  Druten}},\ }\href {\doibase 10.1016/S1049-250X(08)60101-9} {\bibfield
  {journal} {\bibinfo  {journal} {Advances In Atomic, Molecular, and Optical
  Physics}\ }\textbf {\bibinfo {volume} {37}},\ \bibinfo {pages} {181}
  (\bibinfo {year} {1996})}\BibitemShut {NoStop}%
\bibitem [{\citenamefont {Corwin}\ \emph {et~al.}(1999)\citenamefont {Corwin},
  \citenamefont {Kuppens}, \citenamefont {Cho},\ and\ \citenamefont
  {Wieman}}]{CKC99}%
  \BibitemOpen
  \bibfield  {author} {\bibinfo {author} {\bibfnamefont {K.~L.}~\bibnamefont
  {Corwin}}, \bibinfo {author} {\bibfnamefont {S.~J.~M.}~\bibnamefont {Kuppens}},
  \bibinfo {author} {\bibfnamefont {D.}~\bibnamefont {Cho}}, \ and\ \bibinfo
  {author} {\bibfnamefont {C.~E.}\ \bibnamefont {Wieman}},\ }\href {\doibase
  10.1103/PhysRevLett.83.1311} {\bibfield  {journal} {\bibinfo  {journal}
  {Physical Review Letters}\ }\textbf {\bibinfo {volume} {83}},\ \bibinfo
  {pages} {1311} (\bibinfo {year} {1999})}\BibitemShut {NoStop}%
\bibitem [{\citenamefont {Metcalf}\ and\ \citenamefont {van~der
  Straten}(1999)}]{MS99b}%
  \BibitemOpen
  \bibfield  {author} {\bibinfo {author} {\bibfnamefont {H.~J.}\ \bibnamefont
  {Metcalf}}\ and\ \bibinfo {author} {\bibfnamefont {P.}~\bibnamefont {van~der
  Straten}},\ }\href {\doibase 10.1007/978-1-4612-1470-0} {\emph {\bibinfo
  {title} {{Laser Cooling and Trapping}}}},\ Graduate Texts in Contemporary
  Physics\ (\bibinfo  {publisher} {Springer New York},\ \bibinfo {address} {New
  York, NY},\ \bibinfo {year} {1999})\BibitemShut {NoStop}%
\bibitem [{\citenamefont {Townsend}\ \emph {et~al.}(1995)\citenamefont
  {Townsend}, \citenamefont {Edwards}, \citenamefont {Cooper}, \citenamefont
  {Zetie}, \citenamefont {Foot}, \citenamefont {Steane}, \citenamefont
  {Szriftgiser}, \citenamefont {Perrin},\ and\ \citenamefont
  {Dalibard}}]{TEC95}%
  \BibitemOpen
  \bibfield  {author} {\bibinfo {author} {\bibfnamefont {C.~G.}~\bibnamefont
  {Townsend}}, \bibinfo {author} {\bibfnamefont {N.~H.}~\bibnamefont {Edwards}},
  \bibinfo {author} {\bibfnamefont {C.~J.}~\bibnamefont {Cooper}}, \bibinfo
  {author} {\bibfnamefont {K.~P.}~\bibnamefont {Zetie}}, \bibinfo {author}
  {\bibfnamefont {C.~J.}\ \bibnamefont {Foot}}, \bibinfo {author}
  {\bibfnamefont {A.~M.}\ \bibnamefont {Steane}}, \bibinfo {author}
  {\bibfnamefont {P.}~\bibnamefont {Szriftgiser}}, \bibinfo {author}
  {\bibfnamefont {H.}~\bibnamefont {Perrin}}, \ and\ \bibinfo {author}
  {\bibfnamefont {J.}~\bibnamefont {Dalibard}},\ }\href {\doibase
  10.1103/PhysRevA.52.1423} {\bibfield  {journal} {\bibinfo  {journal}
  {Physical Review A}\ }\textbf {\bibinfo {volume} {52}},\ \bibinfo {pages}
  {1423} (\bibinfo {year} {1995})}\BibitemShut {NoStop}%
\bibitem [{\citenamefont {Sesko}\ \emph {et~al.}(1989)\citenamefont {Sesko},
  \citenamefont {Walker}, \citenamefont {Monroe}, \citenamefont {Gallagher},\
  and\ \citenamefont {Wieman}}]{SWM89}%
  \BibitemOpen
  \bibfield  {author} {\bibinfo {author} {\bibfnamefont {D.}~\bibnamefont
  {Sesko}}, \bibinfo {author} {\bibfnamefont {T.}~\bibnamefont {Walker}},
  \bibinfo {author} {\bibfnamefont {C.}~\bibnamefont {Monroe}}, \bibinfo
  {author} {\bibfnamefont {A.}~\bibnamefont {Gallagher}}, \ and\ \bibinfo
  {author} {\bibfnamefont {C.~E.}\ \bibnamefont {Wieman}},\ }\href {\doibase
  10.1103/PhysRevLett.63.961} {\bibfield  {journal} {\bibinfo  {journal}
  {Physical Review Letters}\ }\textbf {\bibinfo {volume} {63}},\ \bibinfo
  {pages} {961} (\bibinfo {year} {1989})}\BibitemShut {NoStop}%
\bibitem [{\citenamefont {Chang}\ \emph {et~al.}(2008)\citenamefont {Chang},
  \citenamefont {Gritsev}, \citenamefont {Morigi}, \citenamefont
  {Vuleti{\'{c}}}, \citenamefont {Lukin},\ and\ \citenamefont
  {Demler}}]{CGM08}%
  \BibitemOpen
  \bibfield  {author} {\bibinfo {author} {\bibfnamefont {D.~E.}\ \bibnamefont
  {Chang}}, \bibinfo {author} {\bibfnamefont {V.}~\bibnamefont {Gritsev}},
  \bibinfo {author} {\bibfnamefont {G.}~\bibnamefont {Morigi}}, \bibinfo
  {author} {\bibfnamefont {V.}~\bibnamefont {Vuleti{\'{c}}}}, \bibinfo {author}
  {\bibfnamefont {M.~D.}\ \bibnamefont {Lukin}}, \ and\ \bibinfo {author}
  {\bibfnamefont {E.~A.}\ \bibnamefont {Demler}},\ }\href {\doibase
  10.1038/nphys1074} {\bibfield  {journal} {\bibinfo  {journal} {Nature
  Physics}\ }\textbf {\bibinfo {volume} {4}},\ \bibinfo {pages} {884} (\bibinfo
  {year} {2008})}\BibitemShut {NoStop}%
\bibitem [{\citenamefont {Naik}\ \emph {et~al.}(2020)\citenamefont {Naik},
  \citenamefont {Eneriz-Imaz}, \citenamefont {Carey}, \citenamefont
  {Freegarde}, \citenamefont {Minardi}, \citenamefont {Battelier},
  \citenamefont {Bouyer},\ and\ \citenamefont {Bertoldi}}]{NEC20}%
  \BibitemOpen
  \bibfield  {author} {\bibinfo {author} {\bibfnamefont {D.~S.}\ \bibnamefont
  {Naik}}, \bibinfo {author} {\bibfnamefont {H.}~\bibnamefont {Eneriz-Imaz}},
  \bibinfo {author} {\bibfnamefont {M.}~\bibnamefont {Carey}}, \bibinfo
  {author} {\bibfnamefont {T.}~\bibnamefont {Freegarde}}, \bibinfo {author}
  {\bibfnamefont {F.}~\bibnamefont {Minardi}}, \bibinfo {author} {\bibfnamefont
  {B.}~\bibnamefont {Battelier}}, \bibinfo {author} {\bibfnamefont
  {P.}~\bibnamefont {Bouyer}}, \ and\ \bibinfo {author} {\bibfnamefont
  {A.}~\bibnamefont {Bertoldi}},\ }\href {\doibase
  10.1103/PhysRevResearch.2.013212} {\bibfield  {journal} {\bibinfo  {journal}
  {Physical Review Research}\ }\textbf {\bibinfo {volume} {2}},\ \bibinfo
  {pages} {013212} (\bibinfo {year} {2020})},\ \Eprint
  {http://arxiv.org/abs/1910.12849} {arXiv:1910.12849} \BibitemShut {NoStop}%
\bibitem [{\citenamefont {Savard}\ \emph {et~al.}(1997)\citenamefont {Savard},
  \citenamefont {O'Hara},\ and\ \citenamefont {Thomas}}]{SOT97}%
  \BibitemOpen
  \bibfield  {author} {\bibinfo {author} {\bibfnamefont {T.~A.}~\bibnamefont
  {Savard}}, \bibinfo {author} {\bibfnamefont {K.~M.}~\bibnamefont {O'Hara}}, \
  and\ \bibinfo {author} {\bibfnamefont {J.~E.}~\bibnamefont {Thomas}},\ }\href
  {\doibase 10.1103/PhysRevA.56.R1095} {\bibfield  {journal} {\bibinfo
  {journal} {Physical Review A}\ }\textbf {\bibinfo {volume} {56}},\ \bibinfo
  {pages} {R1095} (\bibinfo {year} {1997})}\BibitemShut {NoStop}%
\bibitem [{\citenamefont {Friebel}\ \emph {et~al.}(1998)\citenamefont
  {Friebel}, \citenamefont {D'Andrea}, \citenamefont {Walz}, \citenamefont
  {Weitz},\ and\ \citenamefont {H{\"{a}}nsch}}]{FDW98}%
  \BibitemOpen
  \bibfield  {author} {\bibinfo {author} {\bibfnamefont {S.}~\bibnamefont
  {Friebel}}, \bibinfo {author} {\bibfnamefont {C.}~\bibnamefont {D'Andrea}},
  \bibinfo {author} {\bibfnamefont {J.}~\bibnamefont {Walz}}, \bibinfo {author}
  {\bibfnamefont {M.}~\bibnamefont {Weitz}}, \ and\ \bibinfo {author}
  {\bibfnamefont {T.~W.}\ \bibnamefont {H{\"{a}}nsch}},\ }\href {\doibase
  10.1103/PhysRevA.57.R20} {\bibfield  {journal} {\bibinfo  {journal} {Physical
  Review A}\ }\textbf {\bibinfo {volume} {57}},\ \bibinfo {pages} {R20}
  (\bibinfo {year} {1998})}\BibitemShut {NoStop}%
\bibitem [{\citenamefont {Cline}\ \emph {et~al.}(1994)\citenamefont {Cline},
  \citenamefont {Miller}, \citenamefont {Matthews},\ and\ \citenamefont
  {Heinzen}}]{CMM94}%
  \BibitemOpen
  \bibfield  {author} {\bibinfo {author} {\bibfnamefont {R.~A.}\ \bibnamefont
  {Cline}}, \bibinfo {author} {\bibfnamefont {J.~D.}\ \bibnamefont {Miller}},
  \bibinfo {author} {\bibfnamefont {M.~R.}\ \bibnamefont {Matthews}}, \ and\
  \bibinfo {author} {\bibfnamefont {D.~J.}\ \bibnamefont {Heinzen}},\ }\href
  {\doibase 10.1364/OL.19.000207} {\bibfield  {journal} {\bibinfo  {journal}
  {Optics Letters}\ }\textbf {\bibinfo {volume} {19}},\ \bibinfo {pages} {207}
  (\bibinfo {year} {1994})}\BibitemShut {NoStop}%
\bibitem [{\citenamefont {Roati}\ \emph {et~al.}(2001)\citenamefont {Roati},
  \citenamefont {Jastrzebski}, \citenamefont {Simoni}, \citenamefont
  {Modugno},\ and\ \citenamefont {Inguscio}}]{RJS01}%
  \BibitemOpen
  \bibfield  {author} {\bibinfo {author} {\bibfnamefont {G.}~\bibnamefont
  {Roati}}, \bibinfo {author} {\bibfnamefont {W.}~\bibnamefont {Jastrzebski}},
  \bibinfo {author} {\bibfnamefont {A.}~\bibnamefont {Simoni}}, \bibinfo
  {author} {\bibfnamefont {G.}~\bibnamefont {Modugno}}, \ and\ \bibinfo
  {author} {\bibfnamefont {M.}~\bibnamefont {Inguscio}},\ }\href {\doibase
  10.1103/PhysRevA.63.052709} {\bibfield  {journal} {\bibinfo  {journal}
  {Physical Review A}\ }\textbf {\bibinfo {volume} {63}},\ \bibinfo {pages}
  {052709} (\bibinfo {year} {2001})}\BibitemShut {NoStop}%
\bibitem [{\citenamefont {J{\'{a}}uregui}\ \emph {et~al.}(2001)\citenamefont
  {J{\'{a}}uregui}, \citenamefont {Poli}, \citenamefont {Roati},\ and\
  \citenamefont {Modugno}}]{JPR01}%
  \BibitemOpen
  \bibfield  {author} {\bibinfo {author} {\bibfnamefont {R.}~\bibnamefont
  {J{\'{a}}uregui}}, \bibinfo {author} {\bibfnamefont {N.}~\bibnamefont
  {Poli}}, \bibinfo {author} {\bibfnamefont {G.}~\bibnamefont {Roati}}, \ and\
  \bibinfo {author} {\bibfnamefont {G.}~\bibnamefont {Modugno}},\ }\href
  {\doibase 10.1103/PhysRevA.64.033403} {\bibfield  {journal} {\bibinfo
  {journal} {Physical Review A}\ }\textbf {\bibinfo {volume} {64}},\ \bibinfo
  {pages} {033403} (\bibinfo {year} {2001})},\ \Eprint
  {http://arxiv.org/abs/0103046} {arXiv:0103046 [physics]} \BibitemShut
  {NoStop}%
\bibitem [{\citenamefont {Steck}()}]{Steck2015}%
  \BibitemOpen
  \bibfield  {author} {\bibinfo {author} {\bibfnamefont {D.~A.}\ \bibnamefont
  {Steck}},\ }\href {http://steck.us/alkalidata} {\enquote {\bibinfo {title}
  {{Rubidium 87 D Line Data}},}\ \bibinfo {pages} 
	{available online at http://steck.us/alkalidata (revision 2.2.1, 21 November 2019)}}\BibitemShut {NoStop}%
\end{thebibliography}

%

\end{document}